\documentclass[conference]{IEEEtran}

\usepackage{textcomp}
\usepackage{cite}
\usepackage{amsmath,amssymb,amsfonts}
\usepackage[linesnumbered,ruled,vlined]{algorithm2e}

\usepackage{subfig}
\usepackage{graphicx}
\graphicspath{ {./figures/} }

\usepackage{tabularx}
\usepackage{multirow}
\usepackage{hyperref}

\usepackage{pifont}
\newcommand{\cmark}{\ding{51}}

\def\BibTeX{{\rm B\kern-.05em{\sc i\kern-.025em b}\kern-.08em
    T\kern-.1667em\lower.7ex\hbox{E}\kern-.125emX}}

\hypersetup{pdfauthor={Jorge Martin-Perez},pdftitle={dimensioning-v2x-through-forecast-scaling}}

\begin{document}

\title{Dimensioning of V2X Services in 5G Networks through Forecast-based Scaling}

\author{\IEEEauthorblockN{Jorge Mart\'{i}n-P\'{e}rez\IEEEauthorrefmark{1},
Koteswararao Kondepu\IEEEauthorrefmark{2}\IEEEauthorrefmark{5},
Danny De Vleeschauwer\IEEEauthorrefmark{3},
Venkatarami Reddy\IEEEauthorrefmark{4},\\
Carlos Guimar\~{a}es\IEEEauthorrefmark{1},
Andrea Sgambelluri\IEEEauthorrefmark{5},
Luca Valcarenghi\IEEEauthorrefmark{5},
Chrysa Papagianni\IEEEauthorrefmark{6},
Carlos J. Bernardos\IEEEauthorrefmark{1}}\\
\IEEEauthorblockA{
\IEEEauthorrefmark{1}Universidad Carlos III de Madrid, Spain,
\IEEEauthorrefmark{2}Indian Institute of Technology,  Dharwad, India,\\
\IEEEauthorrefmark{3}Nokia Bell Labs, Antwerp, Belgium,
\IEEEauthorrefmark{4}Indian Institute of Technology, Hyderabad, India,\\
\IEEEauthorrefmark{5}Scuola Superiore Sant'Anna, Pisa, Italy,
\IEEEauthorrefmark{6}University of Amsterdam, Netherlands.
}
}

\markboth
{Mart\'{i}n-P\'{e}rez \headeretal: Dimensioning of V2X Services in 5G Networks through Forecast-based Scaling}
{Mart\'{i}n-P\'{e}rez \headeretal: Dimensioning of V2X Services in 5G Networks through Forecast-based Scaling}

\maketitle

\begin{abstract}
With the increasing adoption of intelligent 
transportation systems and the upcoming era 
of autonomous vehicles, vehicular services (such as, remote driving, cooperative awareness, and hazard warning) will face an ever changing and dynamic environment. Traffic flows on the roads is a critical condition for these services and, therefore, it is of paramount importance to forecast how they will evolve over time.
By knowing future events (such as, traffic
jams), vehicular services can be dimensioned in an on-demand fashion in order to minimize Service Level Agreements (SLAs) violations, thus reducing the chances of car accidents.
This research departs from an evaluation of traditional 
time-series techniques with 
recent Machine Learning (ML)-based solutions to forecast 
traffic flows in the roads of
Torino (Italy).
Given the accuracy of the selected forecasting techniques, a forecast-based scaling algorithm is proposed and evaluated over a set of dimensioning experiments of three distinct vehicular 
services with strict latency requirements.
Results show that the proposed scaling algorithm enables resource savings of up to a 5\% at the cost of incurring in an increase of less than 0.4\% of latency violations.

\end{abstract}

\begin{IEEEkeywords}
Vehicle-to-Network, Scaling, Forecasting, Time-series, Machine Learning
\end{IEEEkeywords}

%

\section{Introduction}
\label{sec:introduction}
The 5\textsuperscript{th} generation (5G) of mobile communications extends the general purpose connectivity design of earlier generations to support a wide variety of use cases with a disparate set of requirements and capabilities in a single and shared physical infrastructure. Namely, enhanced Mobile BroadBand (eMBB), Ultra Reliable Low Latency Communications (URLLC) and massive Machine Type Communications (mMTC), identified as the main network services to be supported by 5G systems, have different requirements in terms of end-to-end (E2E) delay, bandwidth, availability, reliability, and device density. Such distinct services will have a significant impact on how operators manage their infrastructure, with 5G networks shifting from the monolithic architectures of previous generations to a highly modular, highly flexible and highly programmable architectures. Network Function Virtualization (NFV) and Software-Defined Networking (SDN), along with the convergence of mobile networks and Edge/Cloud infrastructures, appear as key enablers for the realization of such vision. In doing so, a custom-fit paradigm becomes available where virtual and isolated networks (the so-called network slicing \cite{7926923}) are provided over the same and shared infrastructure and tailored to particular services and their requirements.

First, services and their network slices (hereinafter referred only as services) are orchestrated in an on-demand fashion. This step can be seen as the initial dimensioning of the service and mostly relies on static and pre-defined information. And second, the service must be continuously and dynamically dimensioned to face changes in the demand, avoiding any degradation of the service performance and violation of Service Level Agreements (SLAs). To this end, more traditional scaling approaches, like static or reactive (e.g., threshold-based) solutions, have been leveraged during the service lifespan. However, these are inflexible in timely (re)allocating network resources, incapable of facing unforeseen events, especially when multiple service must coexist. Consequently, they usually result in over-scaling of the network resources and, thus, a non-optimal dimensioning, which reduces the number of services that can be supported at the same time over the same infrastructure.

An optimal dimensioning of coexistent services is then essential to maximize network resources utilization and to reduce the probability of affecting the performance of one another. 
Forecasting appears as a key supporting capability to aid orchestrator and management entities on their decision-making processes by estimating the future demand of running services. In doing so, these entities can take better decisions regarding the (re)allocation or scaling of resources for specific services, as they will be fed not only with the current state of the network and services but also with their forecasted (future) state. Additionally, preemptive dimensioning actions (e.g., scaling in/out or up/down) can be taken to: \textit{(i)} prepare the services to the expected demand beforehand; and \textit{(ii)} accommodate the time required to dimension services, which depending on the action is expected to be in the order of seconds when reconfiguring the underlying network or minutes if Virtual Network Functions (VNFs) need to be scaled or redeployed.

Vehicular-to-Network (V2N) services are one of the most significant transformations of the automobile industry, gaining increased momentum through 5G networks. V2N services are characterized by having a periodic demand variation over time and by requiring process-intensive and low-latency, reliable computing and communication services. However, as they mostly leverage on the Edge~\cite{7488250} as an alternative to Cloud, the lesser available resources must be managed more efficiently. This is where this work aims to contribute, by targeting the dimensioning of V2N services through forecast-based scaling solutions. To do so, it scales the resources allocated to different VNFs, feeding the decision algorithms with road traffic forecasting information. Notwithstanding, the accuracy of the forecasting information affects the correctness of the scaling decisions. As such, several traditional time series and Machine Learning (ML)-based techniques are evaluated in terms of accuracy, ability to forecast different periods in the future, and to cope with changes in the patterns. Based on the outcomes of the former analysis, selected forecasting techniques are leveraged to improve service scaling decisions, in terms of costs savings and fulfillment of their respective SLAs, when applied to distinct V2N services.

The remainder of this paper is organized as follows:
Section~\ref{sec:related} goes over the related work
with respect to forecasting techniques and their application for forecasting road traffic and for network/service dimensioning purposes. 
Section~\ref{sec:techniques} describes
the selected techniques to forecast
road traffic flow, which performance is evaluated over a road traffic dataset. 
Later, Section~\ref{sec:scaling} proposes a forecast-based scaling algorithm, which makes use of the forecasting information to horizontally scale different V2N services. Finally, Section~\ref{sec:conclusions} presents the main conclusions, pointing out
to future research directions.



\section{Background and Related work}
\label{sec:related}

This Section overviews different approaches to implement forecasting techniques, and existing state-of-the-art on their application for road traffic forecasting and network/service dimensioning purposes.

\subsection{Forecasting Techniques}

As stated in~\cite{Brown1963}, the \textit{``every-day life presents countless situations where one must somehow estimate what will happen in the future, as a basis for reaching a decision or taking action''}. Such estimation can also be interpreted as a prediction or forecast.

Traditionally, forecasting techniques resorted to time series methods, such as Error, Trend, Seasonality (ETS), Auto Regressive Integrated Moving Average (ARIMA)~\cite{LEE201166}, and Triple Exponential Smoothing (TES) (i.e., Holt-Winters)~\cite{Brown1963}. These methods usually require a limited number of computational resources and low-energy because they are mainly based on simple analytical formulas~\cite{MarottaJOCN2019}.

With the increasing growth of available data, forecasting can benefit from ML techniques, such as Long Short-Term Memory (LSTM)~\cite{Hochreiter1997} and Recurrent Neural Networks (RNN). In other words, ML is empowering forecasting techniques with the means to implement complex multivariate analysis accounting for different factors that impact a specific phenomenon. However, in contrast to traditional time series techniques, ML-based forecasting requires a large number of resources and energy, especially for training. Therefore, the available computational resources in some network segments might limit their effectiveness. A careful evaluation of the tradeoff between cost and benefits of utilizing traditional time series versus ML techniques must be conducted~\cite{MakridakisPLos2018}\cite{EuCNC2020}, before applying them to any specific scenario.

\subsection{Road traffic forecasting}

Forecasting techniques have been widely used in road traffic scenarios, since they follow a periodic and variable pattern over time. Traditional time series were firstly address to forecast road traffic flows, with methods such as ETS, ARIMA and TES (i.e., Holt-Winters)~\cite{6482260}\cite{MakridakisPLos2018}.

With the emergence of ML, works such as~\cite{6894591} and~\cite{KONG2019460} appeared as the first to apply, respectively, Stacked AutoEncoders (SAEs) and Restricted Boltzmann Machine (RBM) models to forecast road traffic flows. In~\cite{cnn-forecast}, a deep regression model with four layers (including one input, two hidden and one output layer) is used to forecast vehicle flows in a city. Other works rely on the utilization of LSTM~\cite{7874313}\cite{YANG2019320}, Deep Belief Network (DBN)~\cite{s18103459}, Dynamic Fuzzy Neural Networks (D-FNN)~\cite{li2016research}, and Gated Recurrent Units (GRU)~\cite{7804912}, showing promising results on the application of ML-based techniques for road traffic forecasting.

\subsection{Forecasting, and Network and Service Dimensioning}

Forecasting techniques are already used in telecommunication networks to ease and automate tasks related to the lifecycle management of networks and services. As an example, predictive analytics is a key component of the Zero touch network~\&~Service Management~(ZSM) framework envisioned by ETSI~\cite{gszsm}, as an alternative to static rule-based approaches which are inflexible and hard to manage.

In~\cite{8809570}, Deep Artificial Neural Networks are used to forecast network traffic demands of network slices with different behaviors. Similarly, in~\cite{8057230}, a Holt-Winters-based forecasting analyzes and forecasts traffic requests associated to a particular network slice, which is dynamically corrected based on measure deviations. Whilst the former proactively adapts the resources allocated to different services, the latter implements an admission control algorithm to maximize the acceptance ratio of network slice requests. In~\cite{8663796}, LSTM is used by a dynamic bandwidth resource allocation algorithm, aiming to compute the best resource allocation to reduce packet drop probability.

A dynamic dimensioning of the Access and Mobility management Function (AMF), which relies on traffic load forecasting using Deep Neural Network and LSTM, is proposed in \cite{8553653} and \cite{8647590}. In doing so, scaling decisions can be anticipated, avoiding the increase of the attachment time of user equipment and the percentage of rejected requests. A similar solution is also proposed in \cite{8581800} targeting a dynamic and proactive resource allocation to the AMF, where LSTM, CNN, and a combined CNN-LSTM are used to forecast the traffic evolution of a mobile network.




\bigskip

Proponents resulting from the aforementioned work already point out for the potential of forecasting techniques, either traditional time series or ML-based, to provide useful information for network management operations, including the lifecycle management of services. However, to the best of our knowledge, no previous research has tackled the comparison of time series analaysis and ML-based techniques, nor their performance when applied to scaling operations of V2N services with real road traffic traces. Such scenario requires traffic forecasting techniques to \textit{(i)} adapt to changing road traffic conditions (as e.g. the COVID-19 lockdown witnessed in 2020), to \textit{(ii)} scale vehicular services efficiently. This work addresses both challenges and, ultimately, paves the way for a scaling solution for vehicular services, namely V2N services, with strict E2E delay requirements.




\section{Comparison of forecasting techniques}
\label{sec:techniques}

This section provides a brief description of selected forecasting techniques and how \textit{offline}/\textit{online training} can be implemented, followed by their performance using real road traffic traces.


\subsection{Selected Forecasting Techniques}
\label{subsec:selected}
In the scope of this work, distinct time series analysis and ML-based techniques are selected, namely DES and TES time series techniques, and HTM, LSTM, GRU, TCN, and TCNLSTM ML-based techniques.
These are analyzed considering two types of training: \textit{(i)} an \textit{offline training},
in which algorithms learn their parameters in the training set;
and \textit{(ii)} an \textit{online training}, where the parameters
are also updated as the algorithm performs forecastings (see Figure~\ref{fig:online-training}).
To do so, the \textit{online training} uses a moving window
(called online training window -- see Figure~\ref{fig:online-training}) 
comprising the most recent events, which is used to update its parameters before performing
the imminent forecasting.

The next paragraphs provide an explanation of the
selected forecasting techniques:

\begin{figure}[t]
    \centering
        \includegraphics[width=\columnwidth]{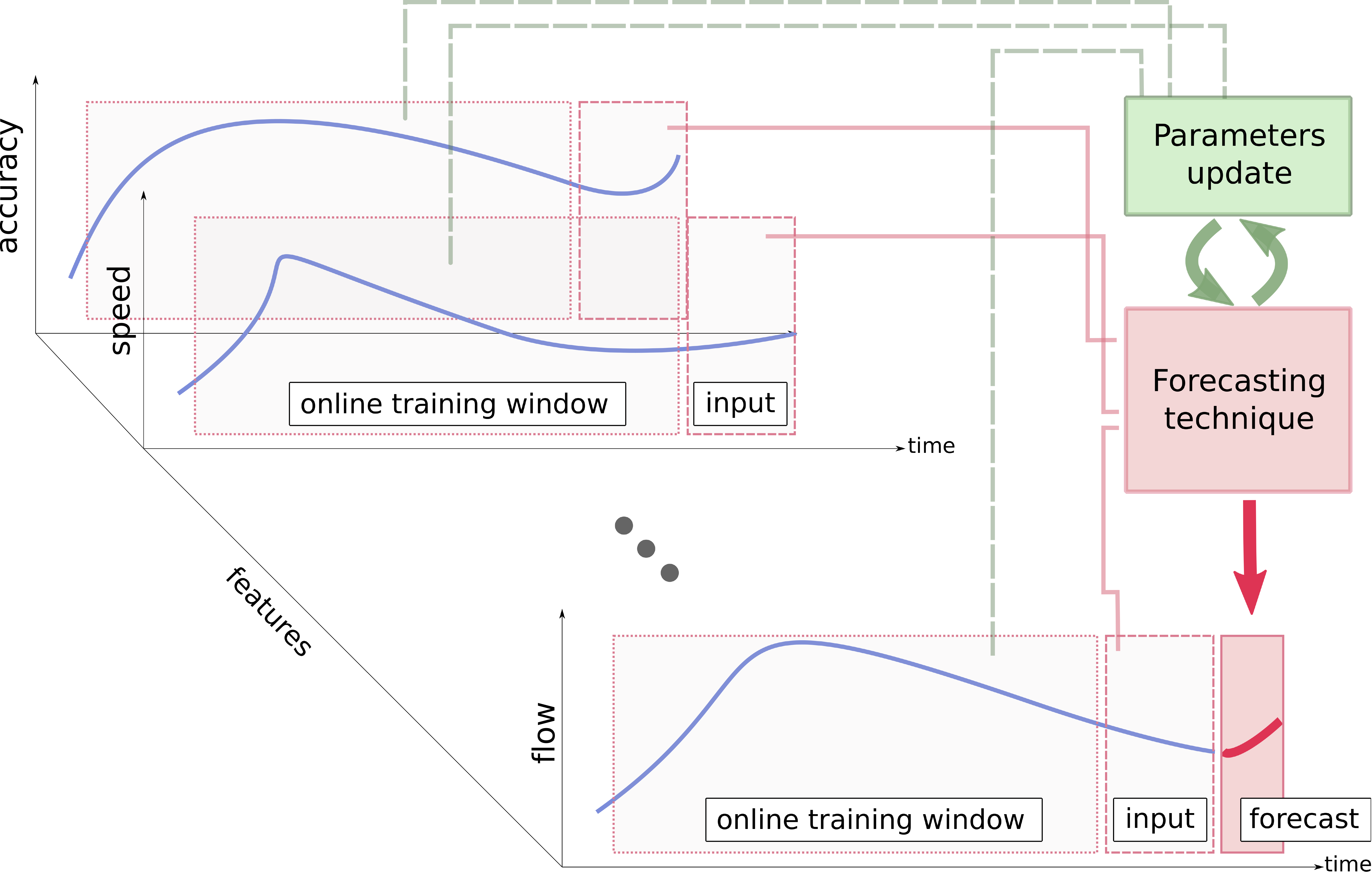}
        
    \caption{\textit{Online training} for traffic forecasting}
 \label{fig:online-training}
\end{figure}

\begin{enumerate}

\item \textbf{Double Exponential Smoothing \text{(DES)}~\cite{Brown1963}:} 
\label{subsec:des}
DES is a forecasting technique based on time series analysis. DES uses a smoothing time scale with a \textit{(i)} \textit{smooth} parameter; and \textit{(ii)} \textit{trend} parameter. The smoothing time scale is obtained based on the previous experienced time interval value of \textit{smooth} and \textit{trend}.
In DES, the \textit{smooth} and \textit{trend} parameters are learned during the \textit{offline training} stage.
If DES is evaluated using \textit{online training}, the \textit{smooth} and \textit{trend} values are also updated using the \textit{online training} window for every forecast.



\item \textbf{Triple Exponential Smoothing \text{(TES)}~\cite{Brown1963}:} 
\label{sec:tes}
TES is another time series analysis technique. 
It exploits three different forecasting parameters, namely \textit{(i)} \textit{smooth}; \textit{(ii)} \textit{trend}; \textit{(iii)} and \textit{seasonality}. 
In TES, the \textit{offline training} is performed by calculating \textit{smooth}, \textit{trend}, and \textit{seasonality} with training set.
Whereas in \textit{online training}, the \textit{smooth}, \textit{trend}, and \textit{seasonality} are updated are
updated for every forecast using the online training window.


\item \textbf{Hierarchical Temporal Memory (HTM)~\cite{Ahmad2017}:}
\label{subsec:html}
The core component of the HTM forecaster is a temporal memory consisting of a two-dimensional array of cells that can either be switched on or off and that evolves with time.
Cells can influence each other via \textit{(i)} \textit{synapses} and \textit{(ii)} \textit{update rules}. 
The offline learning involves adjusting the \textit{synapses} in such a way that the output bit strings resemble the actual input bit strings as much as possible. In that way the temporal memory learns to forecast the next sparse bit strings based on the patterns in the sequence of input bit string it saw. The online learning also updates the \textit{synapses} using the \textit{online training} window.

\item \textbf{Long Short-Term Memory (LSTM)~\cite{Hochreiter1997}:} 
\label{subsec:lstm}
LSTM is a special form of Recurrent Neural Network (RNN) that can learn long-term dependencies based on the information remembered in previous steps of the learning process. It consists of a set of recurrent blocks (i.e., memory blocks), each of the block contains one or more memory cells, and multiplicative units with associated weights, namely, \mbox{\textit{(i)}~\textit{input};} \mbox{\textit{(ii)}~\textit{output}}; and \mbox{\textit{(iii)}~\textit{forget gate}}.
LSTM is one of most successful models for forecasting long-term time series, which can be characterized by different hyper-parameters, specifically the number of hidden layers, the number of neurons, and the batch size.
For the \textit{offline training} approach
neurons' weights are updated
running the back-propagation-through-time~\cite{backpropagation}
over a training dataset.
If LSTM  uses
\textit{online training}, neurons' weights use the
\textit{online training} window to update their values.

%

\item \textbf{Gated Recurrent Unit (GRU)~\cite{gru}:}
\label{subsec:gru}
Gated Recurrent Units (GRUs) are
neurons used in RNNs, and as LSTMs cells, they
store a hidden state that is recurrently fed
into the neuron upon each invocation.
The neuron uses two gates, namely, \textit{(i)} the
\textit{update gate}, and \textit{(ii)} the \textit{reset
gate}.
The former gate is an interpolator between the
previous hidden state, and the candidate
new hidden state; whilst the latter gate
decides what to forget for the new candidate
hidden state.
GRUs keep
track of as much information as possible of
past events. Thus, their use in time-series
forecasting is becoming popular in current
state-of-the-art.
Regarding the offline and online
training, GRU works as the aforementioned LSTM.


\item \textbf{Temporal Convolutional Networks (TCNs)~\cite{cnn-forecast}:}
\label{subsec:temporal-cnn}
TCNs are deep learning architectures based on performing a
temporal convolution over the input. The implemented
version consists of two hidden layers, namely,
\textit{(i)} a first layer to perform the temporal convolution; and
\textit{(ii)} a second layer to re-adjust the dimension of the
convolution output. In particular, the convolution layer
has a window size that is a fourth of the input length in
the time domain. Both the online and \textit{offline training}
update the weights of the densely connected layers, and
follow the same training procedure as the LSTM solution.

\item \textbf{Convolutional LSTM (TCNLSTM)~\cite{Sainath2015}:}
\label{subsec:cnn-lstm}
In the convolutional LSTM, both TCN and LSTM models are combined into a single unified framework. The input features are fed to \textit{(i)} TCN layers; and the \textit{(ii)} LSTM is taking as input the output of the TCN layers to \textit{(iii)} latter feed them into a dense layer. This model is considered to observe the advantage for mapping input features extraction with TCN and interpreting the features with LSTM model. In \cite{Pascanu2014}, it is shown that the LSTM performance can be improved by providing better features. Indeed, TCN helps by reducing the frequency variations in the input features.
TCNLSTM is trained, both in the offline and online procedures,
as the LSTM solution.



\end{enumerate}

\subsection{Performance Evaluation}
\label{sec:flow-prediction}

In order to evaluate the performance of the techniques described above, a real road traffic dataset was collected from 28/01/2020 to 25/03/2020. The dataset comprises measurements from more than 100 road probes in Torino city (Italy), reporting their location, traffic flow, and vehicles speed. This dataset encompasses data pre- and post lockdown due to \textit{COVID-19}.

Each forecasting 
technique is used to forecast the
vehicles/hour
traffic flow $f_t$
seen at Corso Orbassano
road probe\footnote{This is the road probe with highest number of reported measurements in Torino city.} at time $t$.
The set of features
$\phi_i$ at their disposal are those reported by all
road probes $s_j$ ($s_1,\ldots,s_{92}$) of
the dataset.
The numerical value of a feature reported by a probe at instant
$t$ is denoted as $x_t^{\phi_i,s_j}$.
Table~\ref{tab:features} enumerates the
features $\phi_i,\ i=\{1,\ldots,9\}$ used by the selected techniques.
The dataset granularity is of 5~min.,
and throughout this paper $t+1$ represents
the instant $t+5~\text{min}$.

\begin{table}[!t]
\small
    \centering
    \caption{Forecasting features}
    \label{tab:features}
    \begin{tabular}{|p{0.13\columnwidth}|p{0.2\columnwidth}|p{0.2\columnwidth}| p{0.27\columnwidth}|} \hline
     \textbf{Feature} & \textbf{Name} & \textbf{Values} &\textbf{Description} \\\hline
        $\phi_1$ & flow & integer & vehicles/hour\\ \hline
         $\phi_2$ & accuracy & $\{0,\ldots, 100\}$ & percent accuracy of the reported measurement\\ \hline
         $\phi_3$ & speed & float & average vehicles' speed (km/hour)\\ \hline
         $\phi_4$ & distance to Corso Orbassano & [0,35] & distance to Corso Orbassano road probe (km)\\ \hline
         $\phi_5$ & day\_of\_week & $\{1,\ldots,7\}$ & day of the week\\ \hline
         $\phi_6$ & month & $\{1,\ldots, 12\}$ & month of the measurement\\ \hline
         $\phi_7$ & day & $\{1,\ldots,31\}$ & day of the measurement\\ \hline
         $\phi_8$ & year & integer & year of the measurement\\ \hline
         $\phi_9$ & hour\_min & [0,24) & hour+minute/60\\ \hline
    \end{tabular}
\end{table}

Among all analyzed techniques, some of them
can incorporate all features of past events
to forecast the future flow of Corso 
Orbassano road. Thus, they take as input a matrix containing every feature reported by a
road probe
during the last $h$ timestamps:

\begin{equation}
    X_{t,h} = \left(\begin{array}{c c c}
        x_{t-1}^{\phi_1,s_1} & \ldots & x_{t-1}^{\phi_9,s_1}\\
        \vdots & \ddots & \vdots \\
        x_{t-1}^{\phi_1,s_{92}} & \ldots & x_{t-1}^{\phi_9,s_{92}}\\
        \\
        \vdots & \vdots & \vdots\\
        \\
        x_{t-h}^{\phi_1,s_1} & \ldots & x_{t-h}^{\phi_9,s_1}\\
        \vdots & \ddots & \vdots \\
        x_{t-h}^{\phi_1,s_{92}} & \ldots & x_{t-h}^{\phi_9,s_{92}}
    \end{array}\right)
    \label{eq:feat-matrix}
\end{equation}

Since the dataset contains periods of \textit{COVID19} and \textit{non-COVID19},
it is divided in two parts, each with its training and
testing sets, namely:
\begin{itemize}
    \begin{samepage}
    \item \textit{non-COVID-19 scenario:}
        \begin{itemize}
            \item training: 28\textsuperscript{th} January~-~28\textsuperscript{th} February
            \item testing: 29\textsuperscript{th} February~-~07\textsuperscript{th} March
        \end{itemize}
    \item \textit{COVID-19 scenario:}
        \begin{itemize}
            \item training: 06\textsuperscript{th} February~-~07\textsuperscript{th} March
            \item testing: 8\textsuperscript{th} March~-~15\textsuperscript{th} March
        \end{itemize}
    \end{samepage}
\end{itemize}

For the performance evaluation,
the \textit{offline training} occur during the training
sets to learn the techniques' weights/parameters, and the \textit{online
training} updates the learned weights/parameters as the selected
technique forecasts in the testing set.

The selected techniques of Section~\ref{subsec:selected}
were implemented using Python and the TensorFlow library.
LSTM and TCN used the whole feature matrix
$X_{t,h}$ to derive the predictions, whilst the other
techniques just used the traffic flow feature.
Table~\ref{tab:ep} summarizes the parameters that allowed to get the lowest RMSE for each forecasting technique
in the following experiments.


\begin{table}[htb]
\small
    \centering
    \caption{Evaluation Parameters}
    \label{tab:ep}
    \begin{tabular}{|p{0.38\columnwidth}|p{0.27\columnwidth}|p{0.21\columnwidth}|} \hline
     \textbf{Parameter} & \textbf{Forecasting techniques} &\textbf{Value}  \\ \hline
         Level factor ($\alpha$) & DES, TES & 0.5, 0.5\\ \hline
         Trend factor  ($\beta$)  & DES, TES & 0.001, 0.001\\ \hline
         Seasonality factor ($\gamma$) & TES & 0.001 (3 days)\\  \hline
         Hidden layers & TCN, LSTM, GRU, TCNLSTM & 2,2, 1, 4  \\ \hline
         Neurons in hidden layer & TCN, LSTM, GRU, TCNLSTM & 100  \\ \hline
         Epochs   & TCN, LSTM, GRU, TCNLSTM & 100  \\ \hline
        \multirow{2}{*}{history Window size ($h$)} & TCN, LSTM, TCNLSTM & 60~min. \\ \cline{2-3}
        & GRU & 120~min. \\ \hline
         Batch size  & TCN, LSTM, GRU & 5 \\ \hline
         Temporal memory & \multirow{2}{*}{HTM} & 32x2048\\ \cline{1-1}\cline{3-3}
         Encoder representation & & 1024 bit str\\ \hline
    \end{tabular}
\end{table}

\begin{figure*}[ht]%
    \centering
    \subfloat[][\textit{offline training} and non-COVID-19 scenario]{%
        \includegraphics[width=.5\textwidth]{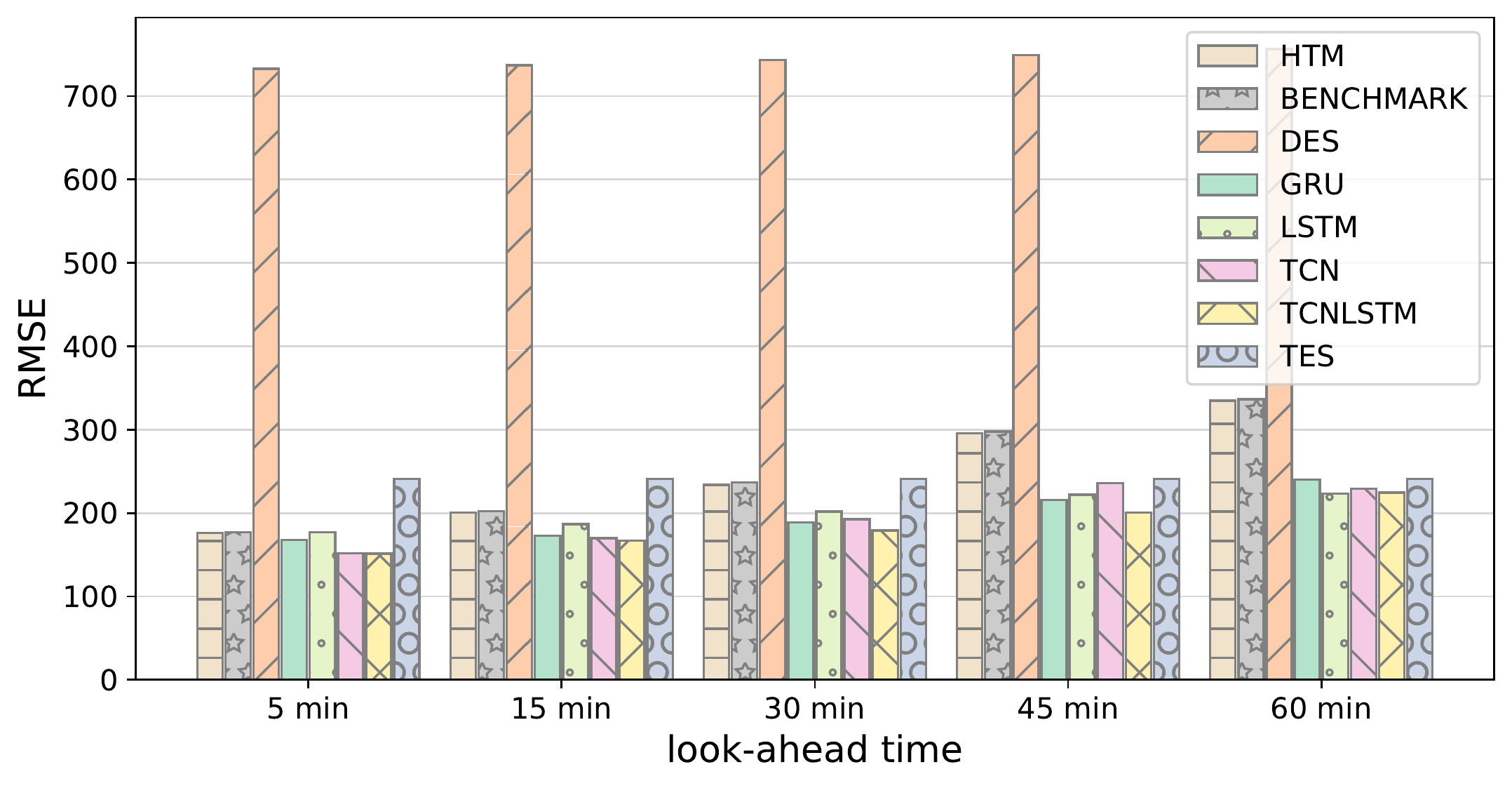}
        \label{fig:offline-non-covid-steps}%
    }%
    \subfloat[][\textit{offline training} and COVID-19 scenario]{%
        \includegraphics[width=.5\textwidth]{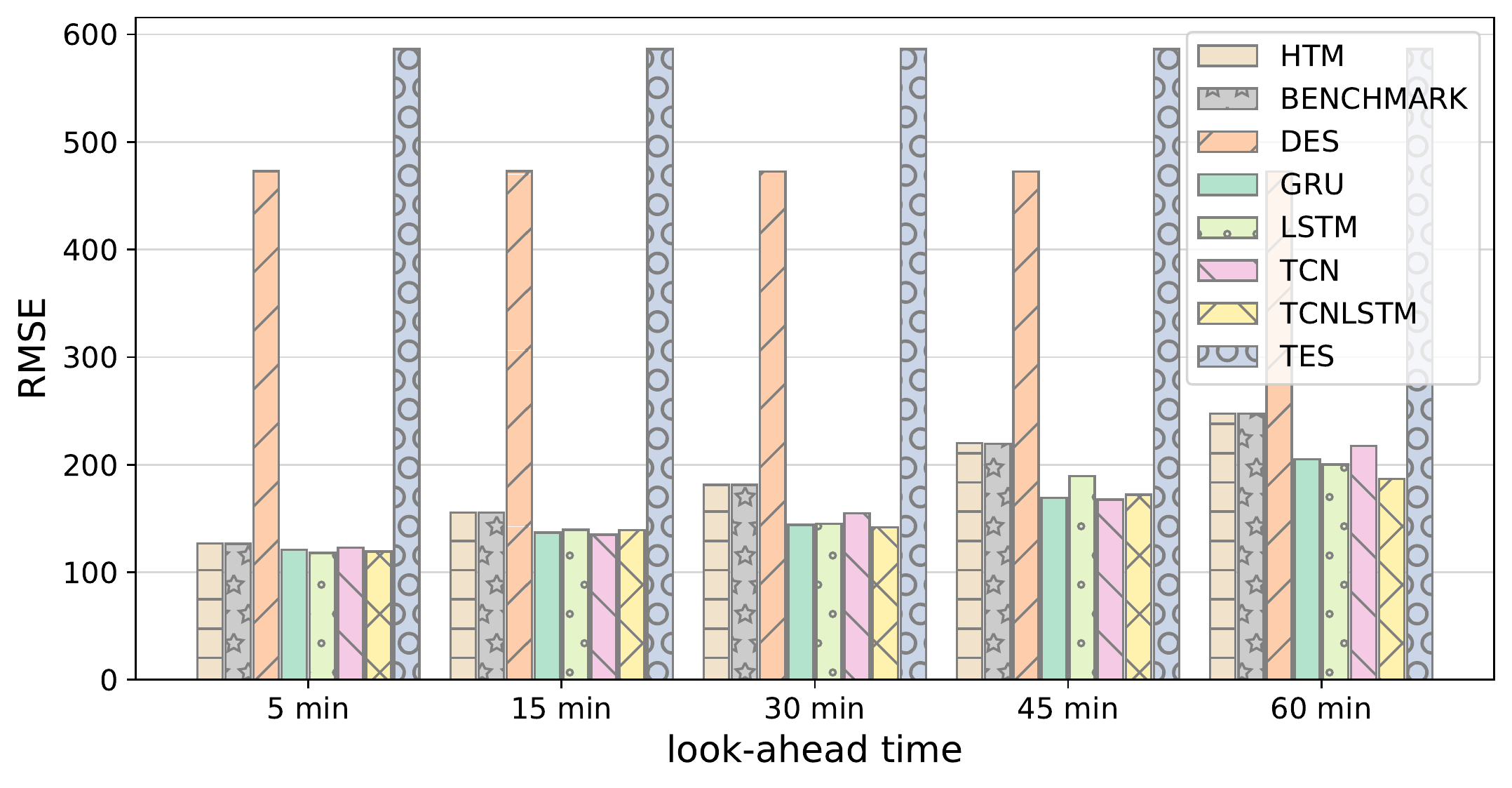}
        \label{fig:offline-covid-steps}%
    }\\
    \subfloat[][\textit{online training} and non-COVID-19 scenario]{%
        \includegraphics[width=.5\textwidth]{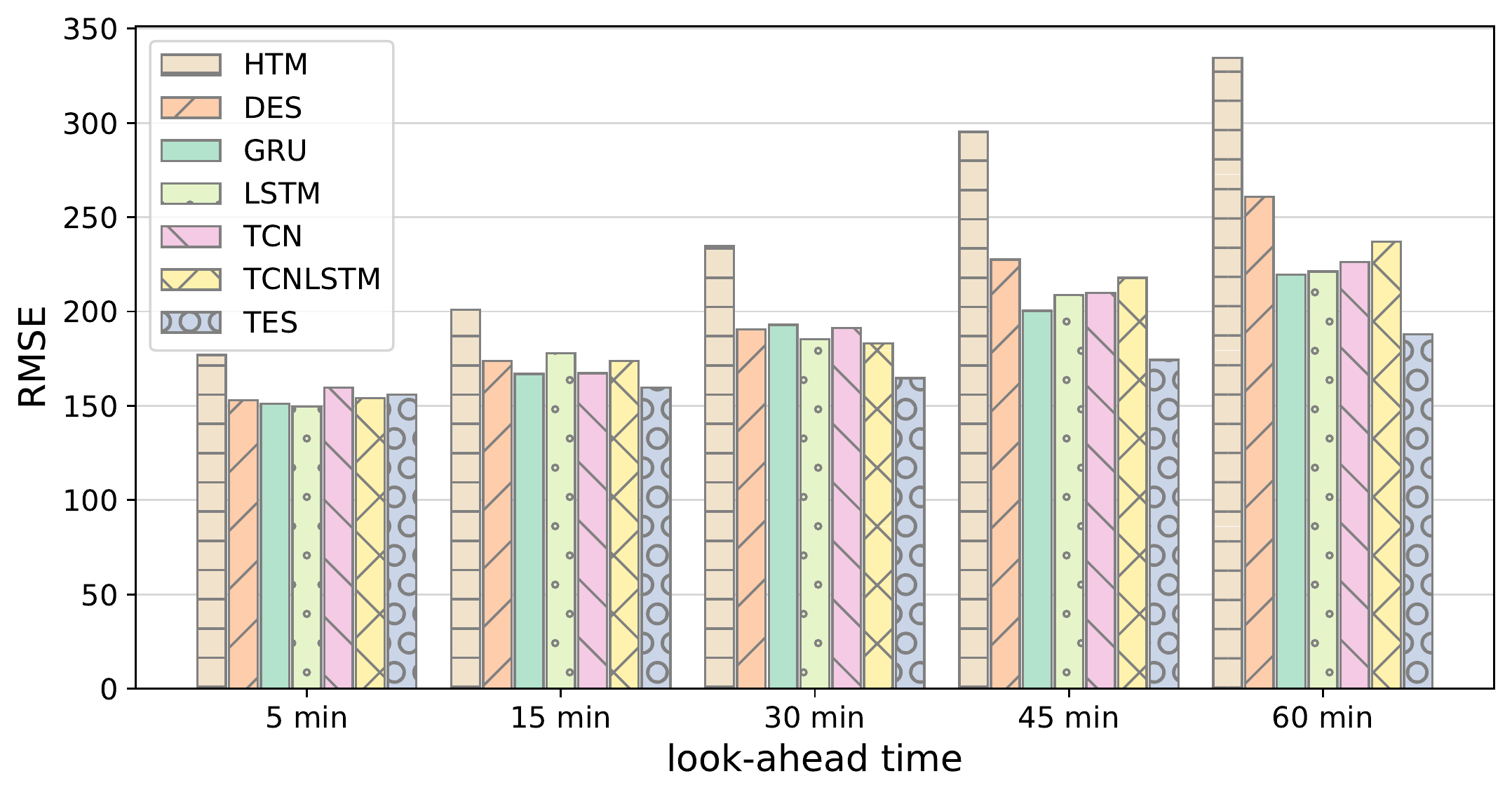}
        \label{fig:online-non-covid-steps}%
    }
    \subfloat[][\textit{online training} and COVID-19 scenario]{%
        \includegraphics[width=.5\textwidth]{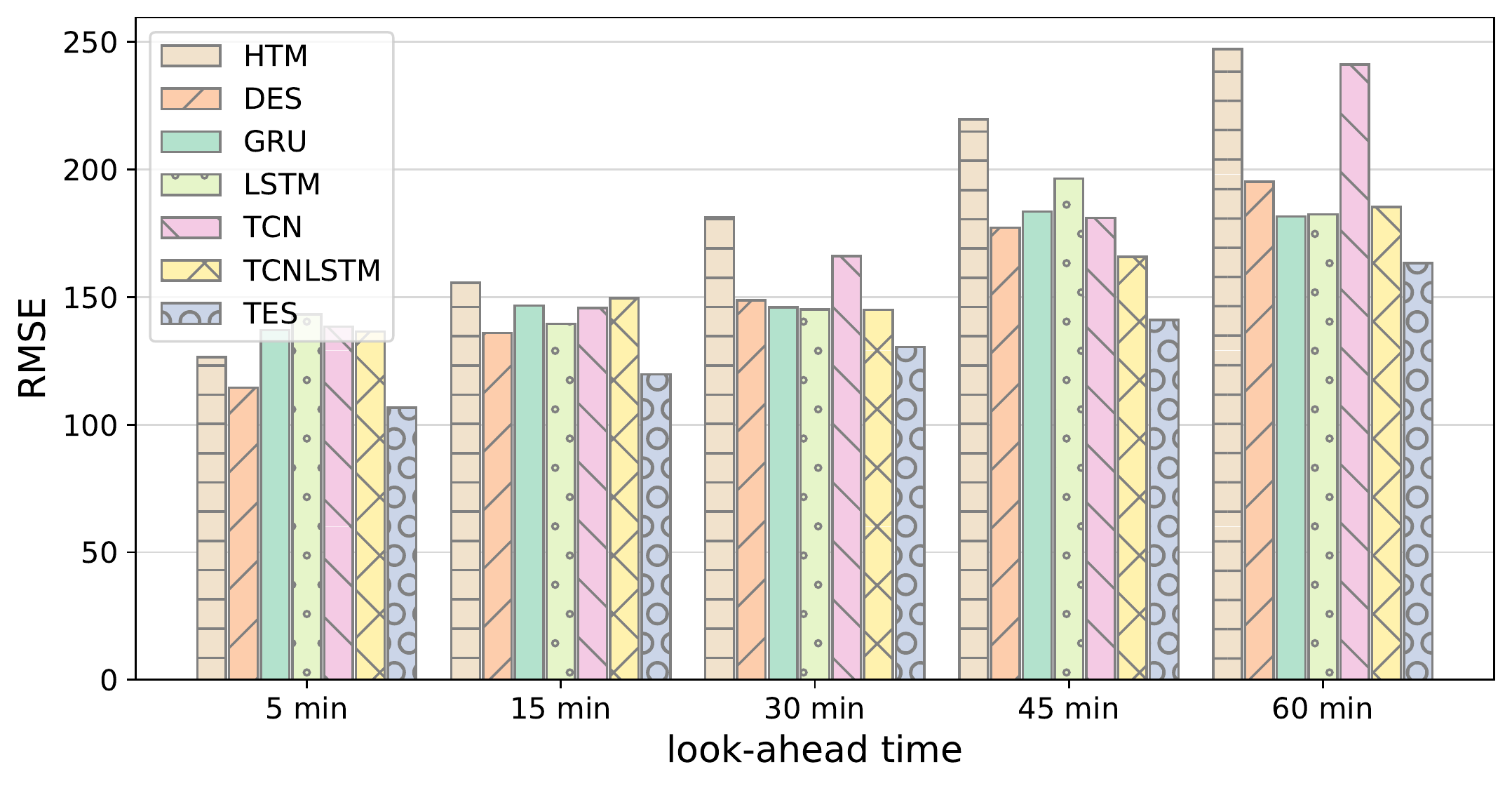}
        \label{fig:online-covid-steps}%
    }%
    \centering
    \caption[]{
        Accuracy of Section~\ref{sec:techniques} look-ahead forecasts.
    }%
    \label{fig:step-ahead-rmse}%
\end{figure*}

\subsubsection{Look-ahead Performance}
\label{subsec:look-ahead-impact}
This Section evaluates the accuracy of the
selected techniques as the look-ahead increases, i.e., as it increases
the number of future traffic flow values to predict.
This analysis is of special importance given the time
that it takes to (de)allocate the resources for a
vehicular service given the applied virtualization
technology, or the type of service.
Results of Figure~\ref{fig:step-ahead-rmse} illustrate how increasing the \textit{look-ahead time} forecast leads to an increasing RMSE in every possible training (i.e., \textit{online} and \textit{offline} training) and dataset combinations (\textit{COVID-19} and \textit{non-COVID-19} scenario), as it becomes more difficult to forecast the traffic further in the future.

Figures~\ref{fig:offline-non-covid-steps} and~\ref{fig:offline-covid-steps} show that the HTM technique did not manage to beat a sample-and-hold
benchmark. Moreover, in the \textit{online training} scenarios, it yielded the worst performance among all analyzed techniques. For the rest of the techniques, the ML-based techniques achieved the best performance for \textit{offline training}. In the \textit{offline training}, DES is not capable of capturing the trend, and the TES pitfalls in the \textit{COVID-19} scenario. Unlike DES and TES, the ML-based techniques can capture the evolving traffic
trend thanks to the update of their hidden
states (except the TCN).
This explains why the ML-based techniques achieve lower RMSE
when using offline forecasting (see
Figure~\ref{fig:offline-non-covid-steps} and
Figure~\ref{fig:offline-covid-steps}).
Figure~\ref{fig:offline-non-covid-steps} and Figure~\ref{fig:offline-covid-steps} show the RMSE values of \textit{offline training} in \textit{non-COVID-19} and \textit{COVID-19} scenarios.
The results presented in Figure~\ref{fig:offline-non-covid-steps} show that DES technique has highest RMSE values, because the smooth and the trend values initially calculated during the training, are not updated in the testing phase.
The other time-series technique (i.e., TES) 
mitigates such problem since its seasonality
factor can capture better the trend.

Figure~\ref{fig:offline-covid-steps} shows the RMSE values of the considered techniques in \textit{offline training} with \textit{COVID-19} traffic. The considered scenario does not show any seasonality during 8\textsuperscript{th} Mar - 15\textsuperscript{th} Mar due to the \textit{COVID-19} lockdown. Thus, the obtained TES results exhibit the highest RMSE value compared to all other techniques. The detailed description about this behavior is discussed later in this section. 

Figure~\ref{fig:online-non-covid-steps} and Figure~\ref{fig:online-covid-steps} show the RMSE values of \textit{online training} in \textit{non-COVID-19} and \textit{COVID-19} scenarios. The TES outperforms all considered ML-based techniques even when the \textit{look-ahead time} increases. In addition, the results show that TES does not increase the RMSE as much as the ML-based techniques. This is due to the fact that it captures faster the new trends of traffic over the time. Thus, the long \textit{look-ahead time} forecasts are better as smoothing, trend, and seasonality are updated for every data point in the test set. Even though the traditional time series techniques (i.e., DES/TES) are limited to uni-variate time series,
the \textit{online} update of their parameters
achieve a better performance than the
ML-based techniques that account for all features
reported in Table~\ref{tab:features}.

\begin{figure}[t]
\vspace{-2mm}
\centering
\includegraphics[width=1.0\columnwidth]{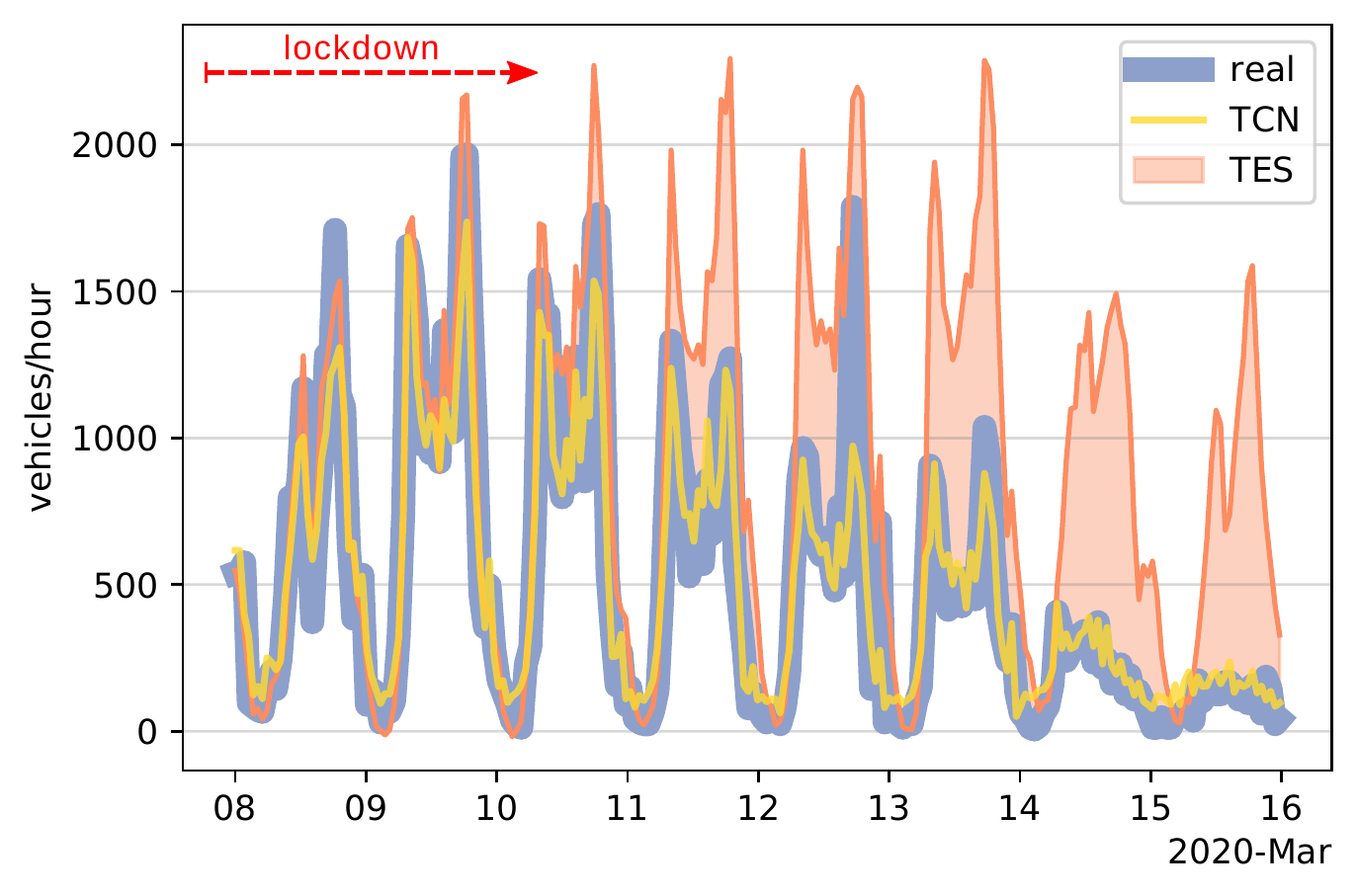}
\vspace{-6mm}
\caption{TES, TCN forecasts vs. real flow values. 5 min. \textit{look-ahead} in \textit{COVID-19} scenario using \textit{offline training.}}
\label{fig:tesbad}
\vspace{-2.5mm}
\end{figure}

\begin{figure*}%
    \centering
    \subfloat[][5 min. \textit{look-ahead}\\ \textit{non-COVID-19}]{%
        \includegraphics[width=.24\textwidth]{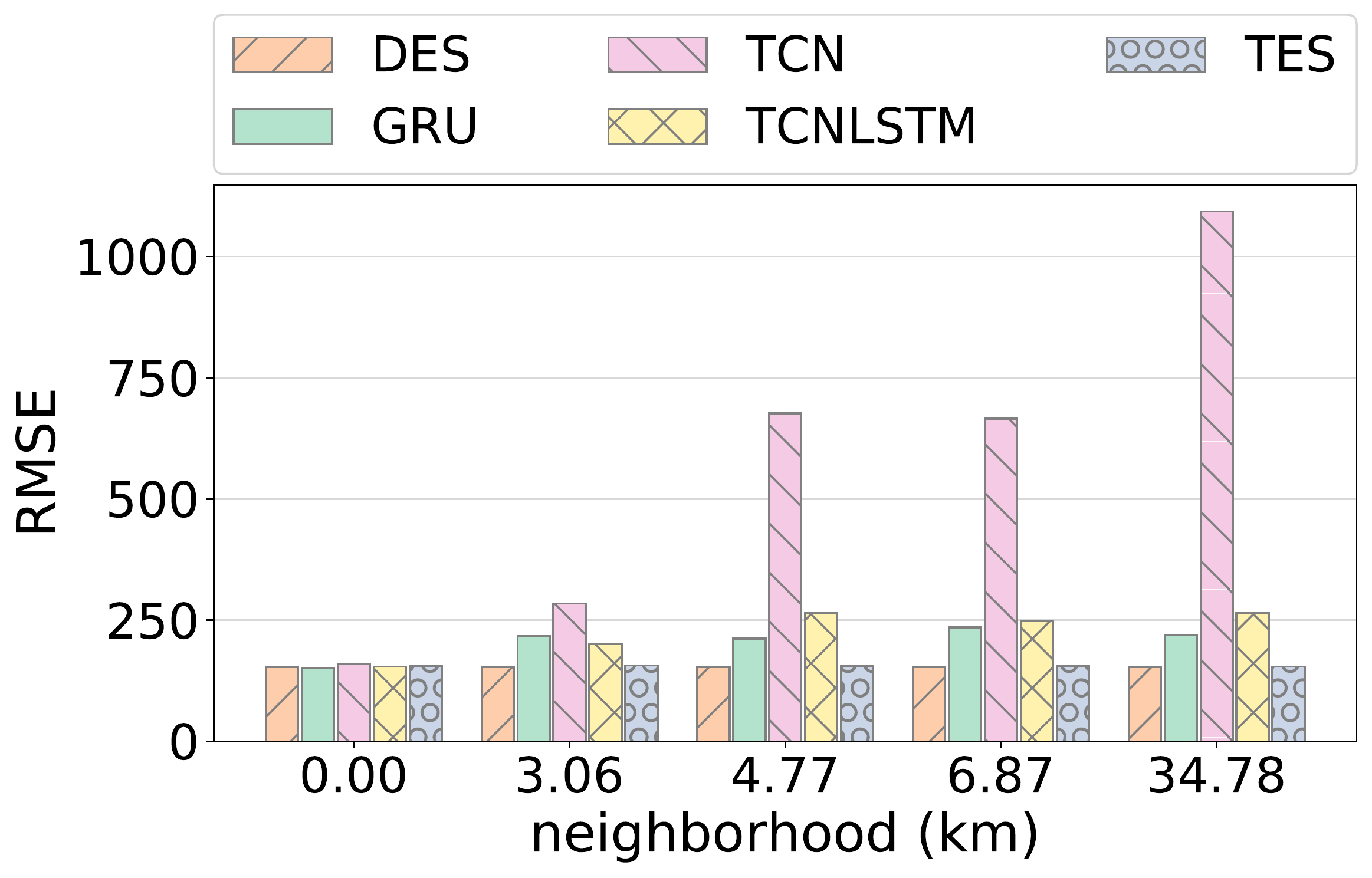}
        \label{fig:non-covid-online-1-step-neigh}%
    }%
    \subfloat[][60 min. \textit{look-ahead}\\ \textit{non-COVID-19}]{
        \includegraphics[width=.24\textwidth]{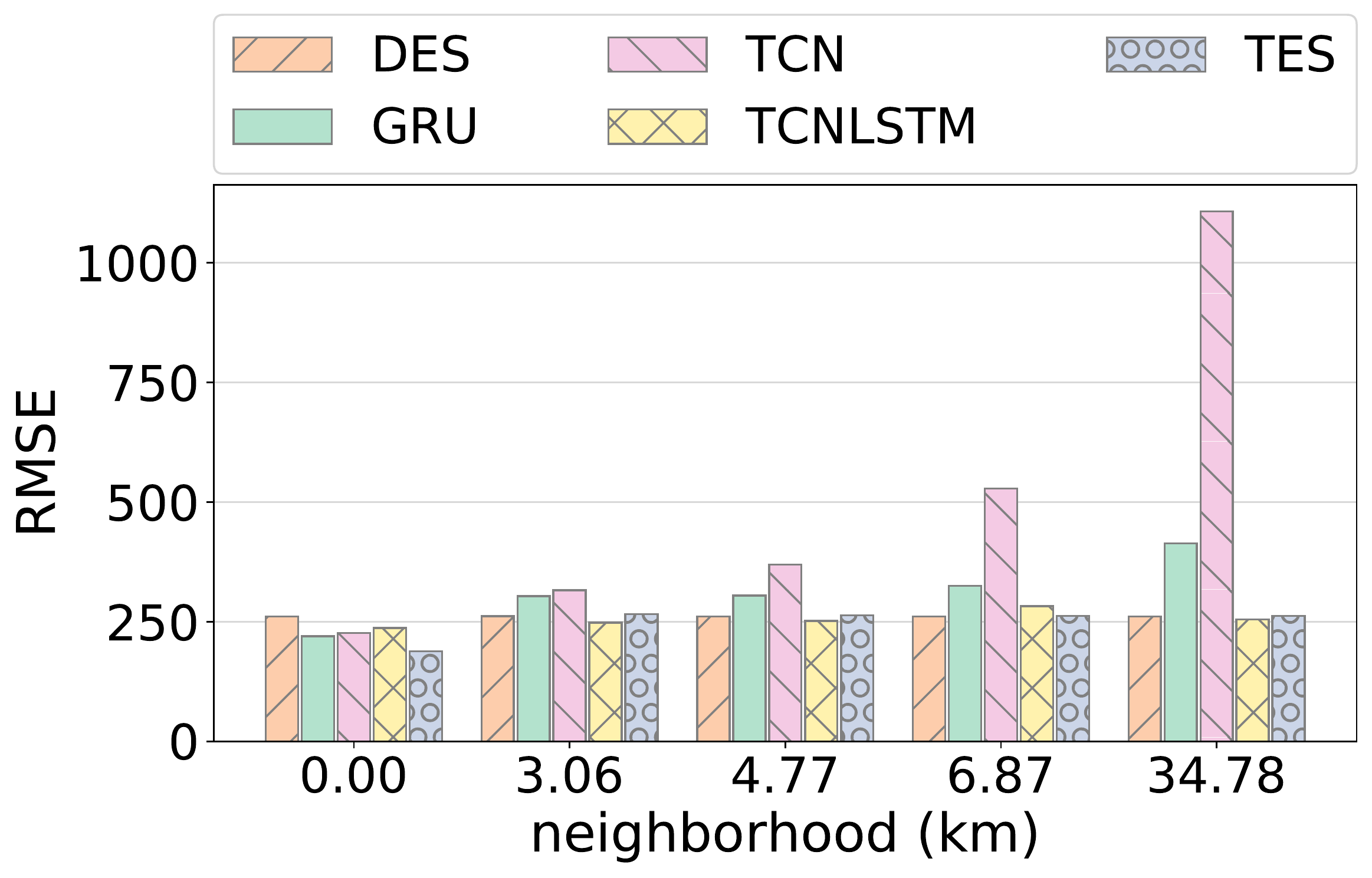}
        \label{fig:non-covid-online-12-step-neigh}%
    }
    \subfloat[][5 min. \textit{look-ahead}\\ \textit{COVID-19}]{%
        \includegraphics[width=.24\textwidth]{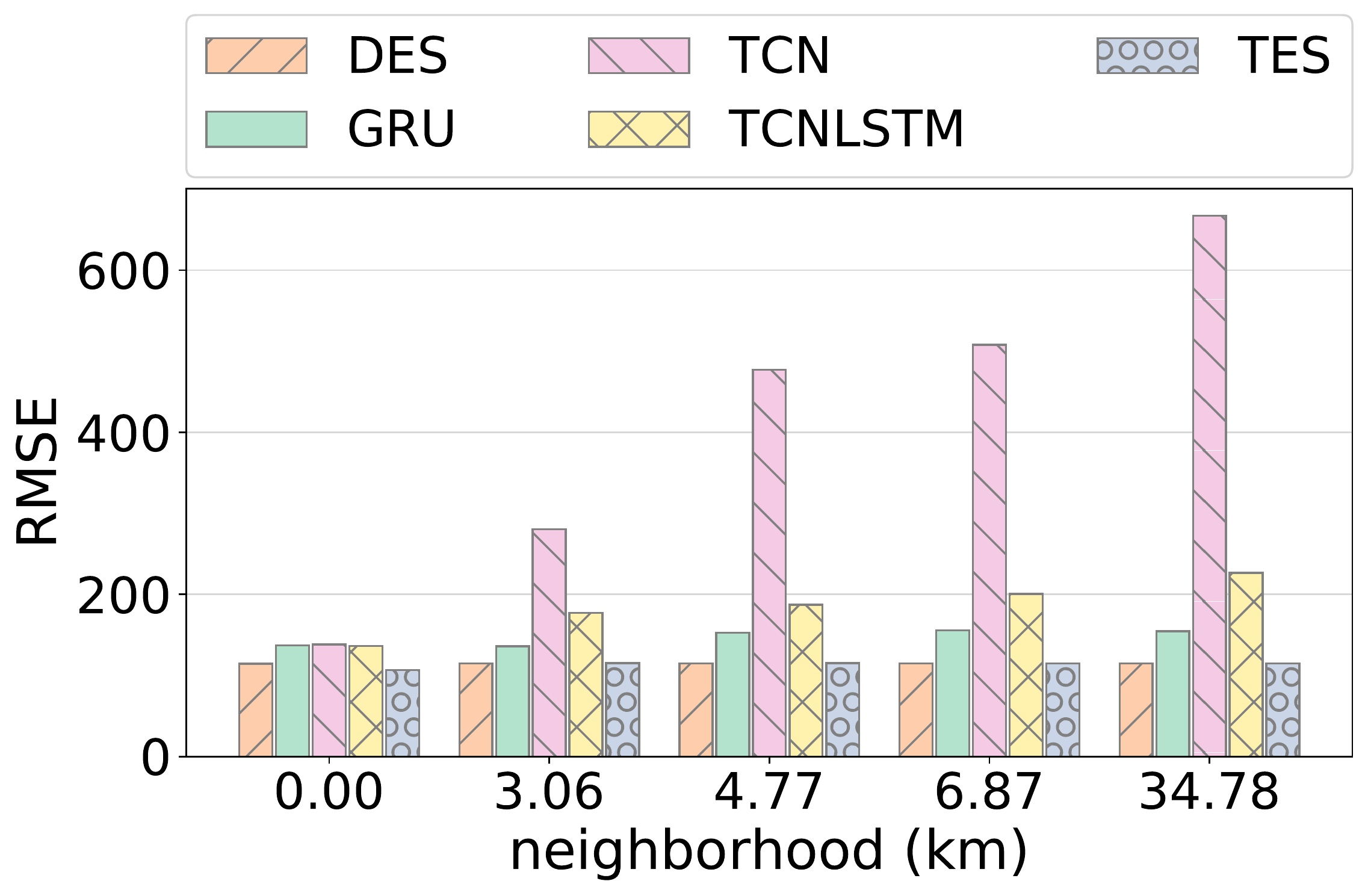}
        \label{fig:covid-online-1-step-neigh}%
    }%
    \subfloat[][60 min. \textit{look-ahead}\\ \textit{COVID-19}]{%
        \includegraphics[width=.24\textwidth]{img/draft-neigh-non-covid-online-steps12.pdf}
        \label{fig:covid-onlin-12-step-neigj}%
    }
    \caption[]{
        Accuracy of 5 min. and 60 min. \textit{look-ahead} forecasting from  Section~\ref{sec:techniques}, using \textit{online training} and varying the neighboring stations.
    }%
    \label{fig:neighbor-increase}%
\end{figure*}

\begin{figure}[!ht]
    \centering
    \includegraphics[width=\columnwidth]{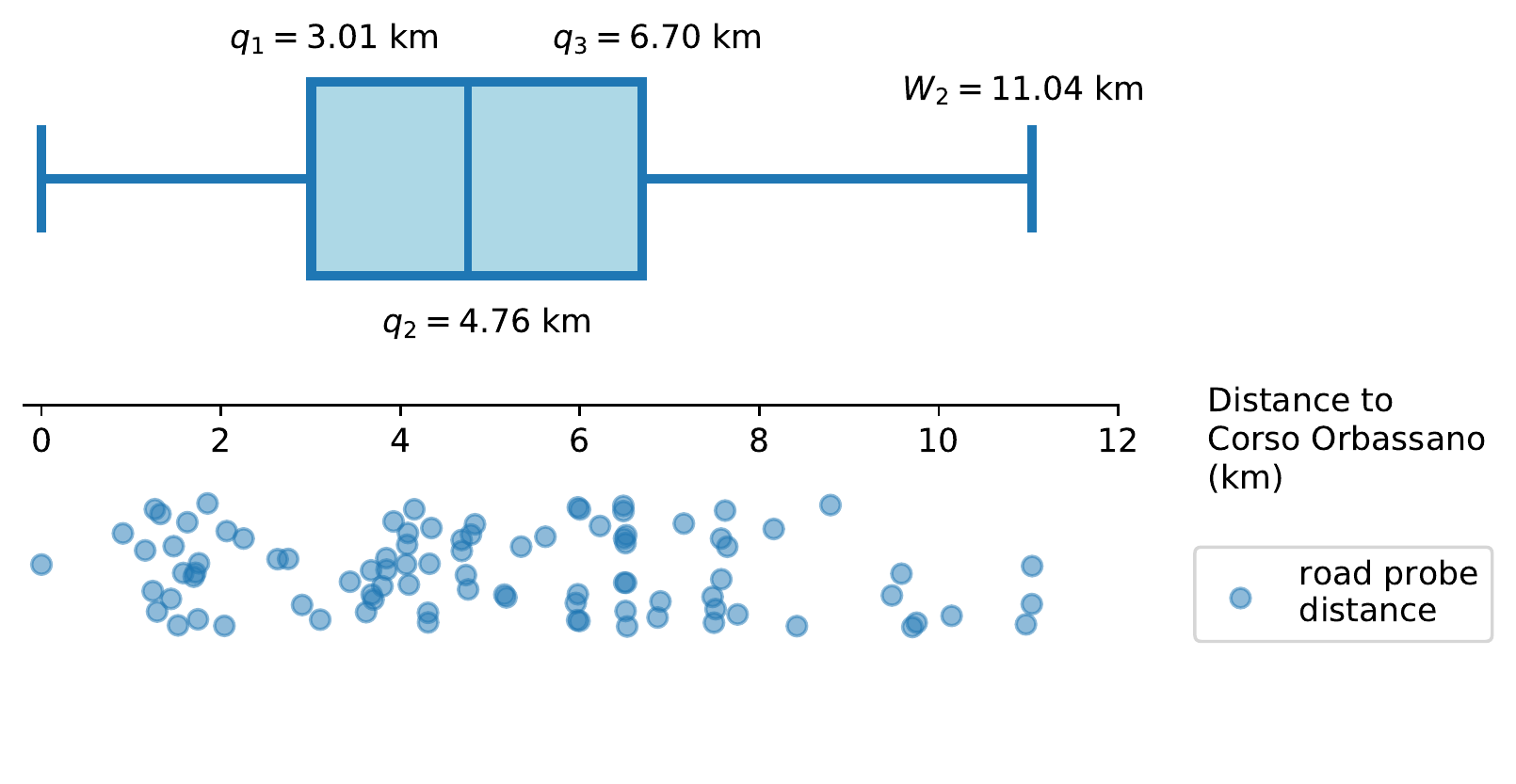}
    \vspace{-10mm}
    \caption{Distances of dataset probes
    towards Corso Orbassano road probe.
    Boxplot above illustrates the quantiles
    for the distances' distribution, along with
    $W_2$, i.e., the last road probe's distance
    that is below $1.5(q_3-q_1)+q_3$.}
    \label{fig:quantiles}
\end{figure}

Finally, Figure~\ref{fig:tesbad} shows the real and the forecasted traffic flow as a function of time. Here, the \textit{look-ahead time} is set to 5 min., and \textit{offline training} is used to forecast the traffic flow during the \textit{COVID-19} scenario (i.e., same conditions as in Figure~\ref{fig:offline-covid-steps}). It is possible to observe that the real traffic flow exhibits a seasonality pattern till 12\textsuperscript{th} Mar. However, latter on traffic flow gradually decreases due to \textit{COVID-19} lockdown. This explains why TES exhibits the highest RMSE values in Figure~\ref{fig:offline-covid-steps}.
ML-based techniques forecast adapts to the
traffic decrease, and among all them,
TCN was selected to show that it forecasts 
traffic flow better than TES.
Because every technique uses \textit{offline training},
TES keeps using the seasonality learned during
the training phase, and it forecasts high
traffic flows even after the decrease.

\subsubsection{Neighborhood performance}
\label{subsec:neighborhood-impact}

Results of Figure~\ref{fig:neighbor-increase}
show whether incorporating the information of
neighboring road probes benefits Corso Orbassano
traffic flow forecast.
Figures~\ref{fig:non-covid-online-1-step-neigh} and Figure~\ref{fig:non-covid-online-12-step-neigh}
show the impact of the neighborhood size to
the RMSE using \textit{online training} in
the \textit{non-COVID-19} scenario.
The experiment considered 5 min. and 60 min. 
\textit{look-ahead time} values, and quantiles in
Figure~\ref{fig:quantiles} as neighborhood
distances.

The increase of the neighborhood leads to
a growth of the training data, due to the
additional information of the
neighboring road probes. As shown in
Figures~\ref{fig:non-covid-online-1-step-neigh}-\ref{fig:covid-onlin-12-step-neigj}, no technique is capable of reducing
the RMSE by having additional neighboring
information.
Among all of them, DES, TES and GRU do not
decrease significantly their performance.
But TCN does, since it convolves every
feature present in the input matrix $X_{t,h}$,
including also the distance to Corso Orbassano
feature. By convolving such feature over
the time domain, the NN cannot distinguish
whether the input corresponds to a Corso
Orbassano measurement or not.
This phenomenon prevents the TCN connections
from giving less relevance to non-correlated
measurements of irrelevant neighboring road
probes.
Note that HTM and LSTM results are not included,
as HTM proprietary implementation could not receive
multiple probes' flows as input, and
both LSTM and GRU are achieving close performance results as shown in Figure~\ref{fig:step-ahead-rmse}. Most of the techniques discussed in this paper should be able to achieve the same RSME value when they take the values of more stations into account than just the Corso Orbassano station. Each technique can set the weights associated to the stations other than Corso Orbassano to 0, washing out the influence of those additional stations completely. The fact that the training does not reach this situation (where all weights associated to stations other than Corso Orbassano are set to 0) means that the training algorithm converges to local minimum rather than the global minimum, thus improvement of the training algorithm is possible.


\section{Forecast-based Scaling for V2N Services}
\label{sec:scaling}
This section devises a queuing theory-based
scaling algorithm that leverages
on the most accurate forecasting techniques of
Section~\ref{sec:flow-prediction}.

Based on the
forecasted flow of future cars,
the scaling algorithm assigns enough resources
to meet the vehicular service
latency requirements. In particular, the performance of
the proposed scaling algorithm is evaluated on
three different V2N services, namely the
\textit{(i)} remote driving; \textit{(ii)} hazard warning services; \textit{(iii)} cooperative awareness, with E2E latency
requirements of 5, 10 and 100~ms, respectively.

\subsection{Forecast-based scaling algorithm}
\label{subsec:proposed-scaling}
To relate number of required resources with the quality of service, a queuing model is utilized. The cars represent the clients while the available automotive service instances represent the servers. A similarity between the handover process and service request is considered. It is assumed that when cars enter in the crossing area, they are requesting the service, as if mobile users handover into another cell. 
For modeling the arrival process, it is necessary to select the arrival process of the cars into the service area. Previous studies have been conducted that provide careful models for automotive traffic~\cite{Gramaglia_comm_lett_2014}.
Similarly, the service time can be modeled as the residence time in a cell of a mobile user. Previous studies provided some models as reported, for example, in~\cite{Courtis_Elec_lett_2002}.
In this work, the proposed model is based on the $M/M/c$ queue, that is at the basis of circuit switching in communications networks~\cite{bertsekas_data_networks}.

Thus, cars are assumed to enter in the coverage area of the service with a Poisson process with arrival rate $\lambda$ and their residence time in the cell is exponentially distributed with average $\frac{1}{\mu}$. Consequently, the average time for which each vehicle spends in the system (i.e., waiting to receive the service and in the service) can be written as:



\begin{equation}
T=\frac{1}{\mu}+\frac{P_Q}{c\mu-\lambda},
\label{eq:avg-service-time}
\end{equation}
where $c$ is the number of available servers and $P_Q$ is the probability that an arrival finds all the services busy.
The expression of $P_Q$ is provided by the Erlang C formula:

\begin{equation}
    P_Q=\frac{p_0 (c\rho)^c}{c! (1-\rho)},
\end{equation}
where $\rho=\frac{\lambda}{c\mu}$ and the probability $p_0$ of having zero clients in the system is:

\begin{equation}
p_0=\left [ \sum_{n=0}^{c-1}\frac{(c \rho)^n}{n!}+ \frac{(c\rho)^c}{c! (1-\rho)} \right ]^{-1}.
\end{equation}

Given the road traffic dataset's flow rate of vehicles $\lambda=f_t$,
and the latency specification $T_0$ of the
V2N services referred in the introduction
of this section (hazard warning, cooperative
awareness, and remote driving),
it is possible to derive the required number
of virtualized service instances $c$ to satisfy the average
E2E latency ($T_0=5$~ms in the case
of remote driving).
More specifically, given the tuple
$(\lambda,\mu)$, $c$ is increased until
the average delay formula (\ref{eq:avg-service-time})
reports a value of $T\le T_0$.
This is the approach used in the proposed
Algorithm~\ref{alg:n-min-scaling} to
derive the required number of servers $c$ and
horizontally scale the V2N service. In particular,
Algorithm~\ref{alg:n-min-scaling} follows a
\mbox{$n$-min.} horizontal scaling approach that
\textit{(i)} forecasts the traffic flow for the next
$n$ minutes; and \textit{(ii)} scales up/down the number
of servers $c$ to meet the average delay for the
maximum flow forecasted within the next $n$ minutes.


%



\subsection{Forecast-based Scaling Performance}
\label{subsec:performance-scaling}
Given the queuing theory framework, this
Section studies the performance of
Algorithm~\ref{alg:n-min-scaling}
to scale remote driving, cooperative awareness, and
hazard warning V2N services.

The performance evaluation is assessed by means of
cost savings, and E2E delay violations. Moreover,
Algorithm~\ref{alg:n-min-scaling} is compared against
two other scaling strategies described latter in this
Section, and during the experiments it used the
most accurate forecasting technique among the ones
evaluated in Section~\ref{sec:flow-prediction}. Results
are derived using \textit{(i)} the aforementioned road traffic dataset;
and \textit{(ii)} reference service rate values of a European
project testbed, namely from \mbox{5G-TRANSFORMER}.

In particular, the service rate $\mu$ is obtained from
\mbox{5G-TRANSFORMER}~\cite{5gt-d5-4}, which reports the results of
what is called an Enhanced Vehicular Service
(EVS), that is, a service that deploys
sensoring, video streaming, and processing
facilities in the edge.
The deliverable reports not
only the required physical resources to
deploy an EVS service, but as well the
flow of cars used to perform their
evaluations.
Moreover, it details that an EVS instance,
i.e.\ $c=1$ in our notation, 
offers a service rate of 
$\mu_{EVS}~=~208.37$~\mbox{vehicles/second}.

\begin{algorithm}[t]
\SetAlgoLined
\KwData{$n, \mu, T_0$}
 \For{$t\in \left\{i\cdot n / 5 min.: i>0\right\}$}{
    $\widehat{f}_{t+n-1}, \ldots, \widehat{f}_{t+1} = $ forecast($X_{t,h}$)\;\label{alg:forecast}
    $\widehat{F} = max \left\{ \widehat{f}_{t+k} \right\}_{k=1}^{n-1}$\;
    $c=1$\;
    $\rho = \widehat{F} / c\mu$\;
    \While{$\frac{1}{\mu} + \frac{P_Q}{c \mu - \widehat{F}} > T_0$}{
        $c = c + 1$\;
    }
    
    scale($c$)\;
 }
 \caption{$n$-min. horizontal scaling}
 \label{alg:n-min-scaling}
\end{algorithm}

\begin{table}[t]
    \caption{
    Best Traffic Flow Forecasting Techniques
    \label{tab:best-technique}
    } 
    \scriptsize
    \begin{tabularx}{\columnwidth}{ |X|X||X|X| }
      \hline
      \multicolumn{2}{|c||}{\textbf{Forecasting task}} & \multicolumn{2}{c|}{\textbf{Best solution}} \\ \hline
      \textbf{Step-ahead} &
      \textbf{Scenario} &
      \textbf{Technique} & 
      \textbf{Online}\\
      \hline
      \multirow{2}{*}{5~min} & \textit{non-COVID-19}  & LSTM & \cmark \\ \cline{2-4}
                             & \textit{COVID-19}  & TES & \cmark \\ \hline
      \multirow{2}{*}{15~min} & \textit{non-COVID-19} &  TES & \cmark \\ \cline{2-4}
                              & \textit{COVID-19} & TES & \cmark \\ \hline
      \multirow{2}{*}{30~min} & \textit{non-COVID-19} & TES & \cmark \\ \cline{2-4}
    & COVID-19  & TES & \cmark \\ \hline
      \multirow{2}{*}{45~min} & \textit{non-COVID-19}  & TES & \cmark \\  \cline{2-4}
                              & \textit{COVID-19}  & TES & \cmark \\  \hline
      \multirow{2}{*}{60~min} & \textit{non-COVID-19}  & TES & \cmark \\  \cline{2-4}
                              & \textit{COVID-19}  & TCNLSTM & \cmark \\  \hline
    \end{tabularx}
\end{table}

The experiments consist in running the
proposed scaling Algorithm~\ref{alg:n-min-scaling}
in the \textit{COVID-19} scenario.
In particular, Algorithm~\ref{alg:n-min-scaling}
has to decide
what is the required number of servers $c$ to
meet the V2N service latency constraint $T_0$
for the forecasted traffic flow (see Algorithm~\ref{alg:n-min-scaling} line~\ref{alg:forecast})
within the next $n$
minutes. The value of $\mu$ is set to be proportional
to $\mu_{EVS}$ depending on the studied V2N service,
and the traffic flow forecasting is done using the
technique that gave the lowest RMSE for
$n$ minutes look-ahead predictions (see Table~\ref{tab:best-technique}).
Additionally, every experiment compares the 
performance of the $n$-min. scaling
Algorithm~\ref{alg:n-min-scaling} with an
average and maximum scaling technique.
Hence, three different scaling strategies are 
considered for evaluation:
\begin{itemize}
    \begin{samepage}
    \item \textit{over-provisioning/max.scaling}: this strategy assumes that the V2N
    service is deployed with $c$ instances
    capable of meeting the average E2E
    delay during peak hours of traffic;
    \item \textit{avg. scaling}: the network
    dimensions the V2N service so that the $c$ 
    instances meet latency restrictions
    considering an average flow of vehicles; and
    \item \textit{$n$-min. scaling}: using the
    best $n$-minutes ahead forecasting
    technique of Table~\ref{tab:best-technique},
    the service is scaled to satisfy the
    peak of traffic forecasted for the next
    $n$ minutes (see Algorithm~\ref{alg:n-min-scaling}).
    \end{samepage}
\end{itemize}

\begin{figure*}[t]%
    \centering
    \subfloat[][Remote driving scaling savings]{%
        \includegraphics[width=.3\textwidth]{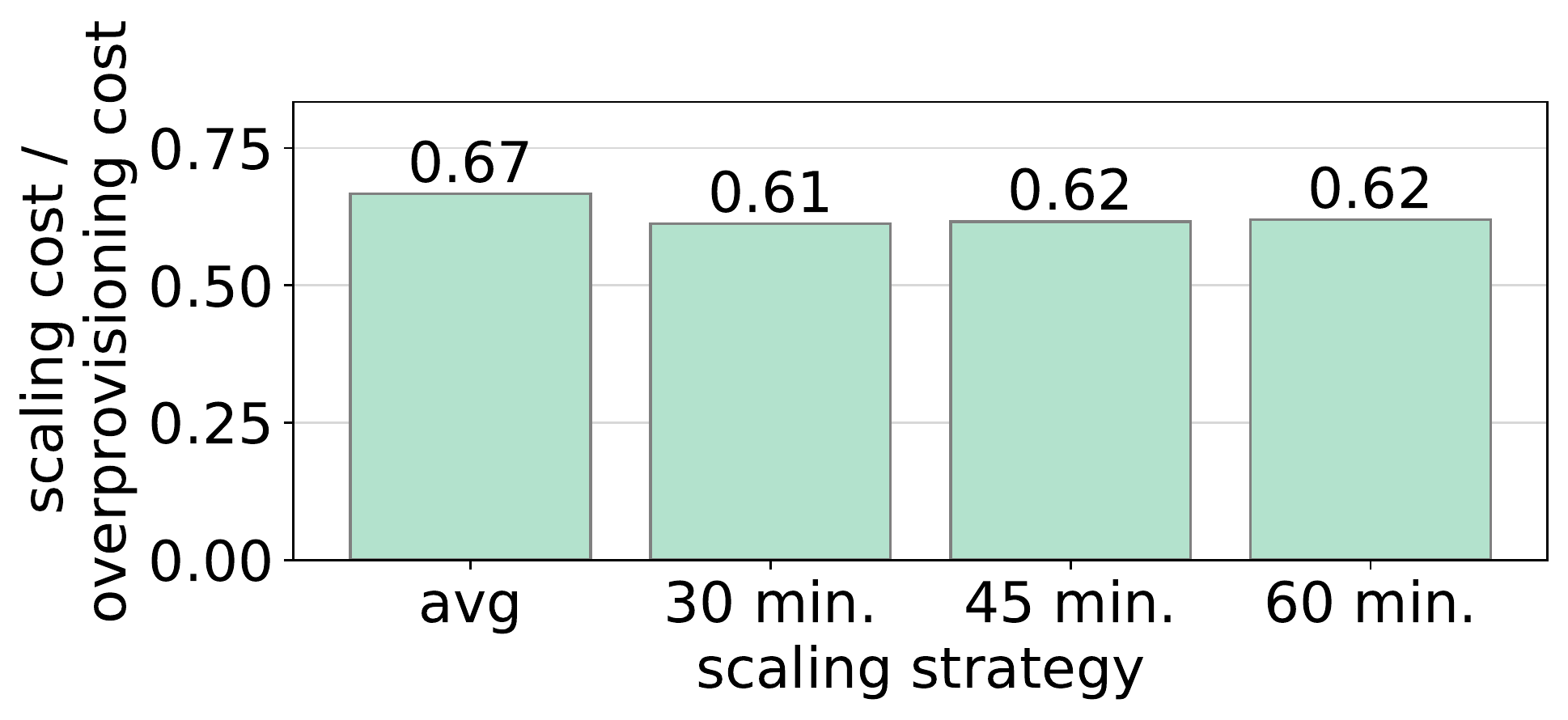}
        \label{fig:cost-remote}%
    }%
    \subfloat[][Cooperative awareness scaling savings]{%
        \includegraphics[width=.3\textwidth]{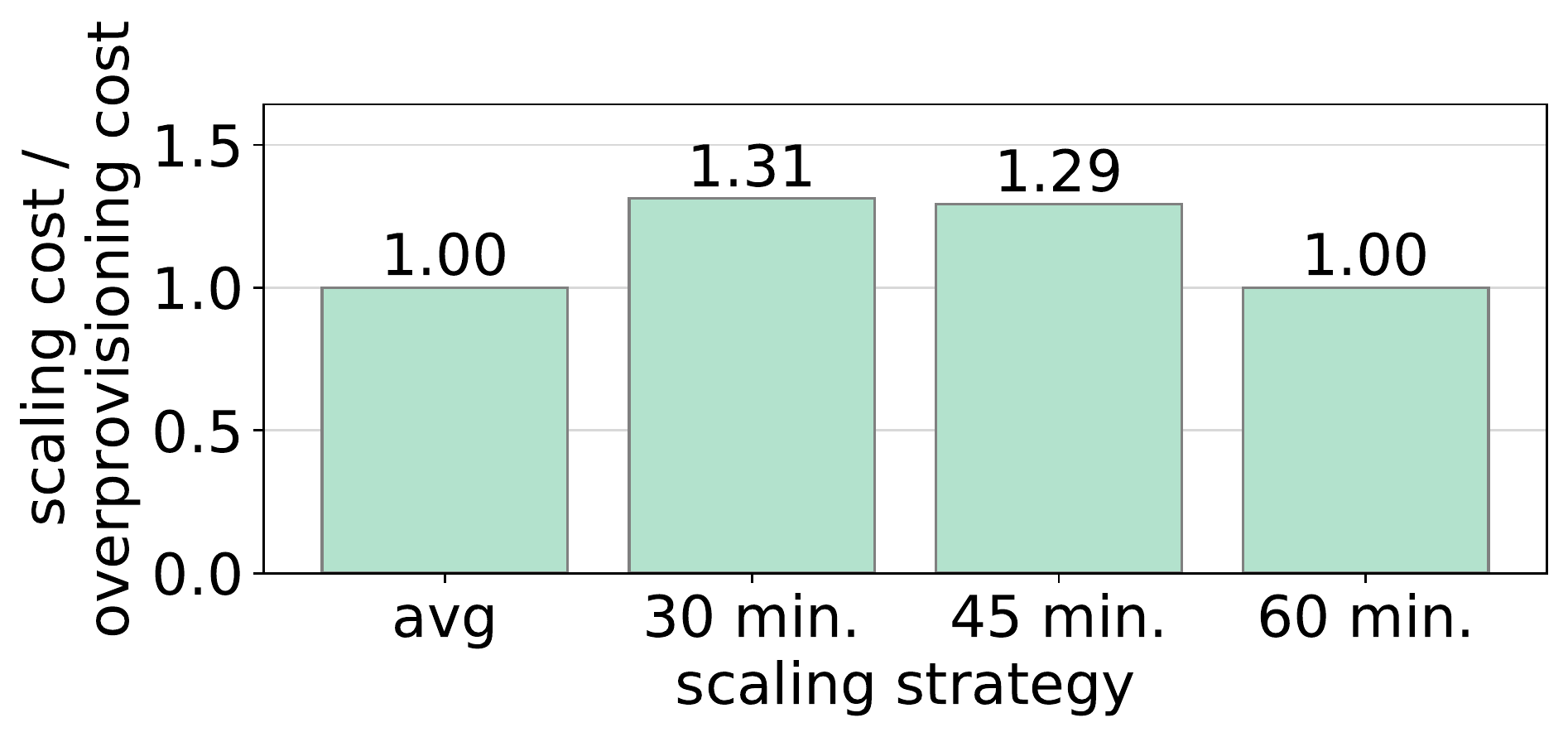}
        \label{fig:cost-cooperative}%
    }
    \subfloat[][Hazard warning scaling savings]{%
        \includegraphics[width=.3\textwidth]{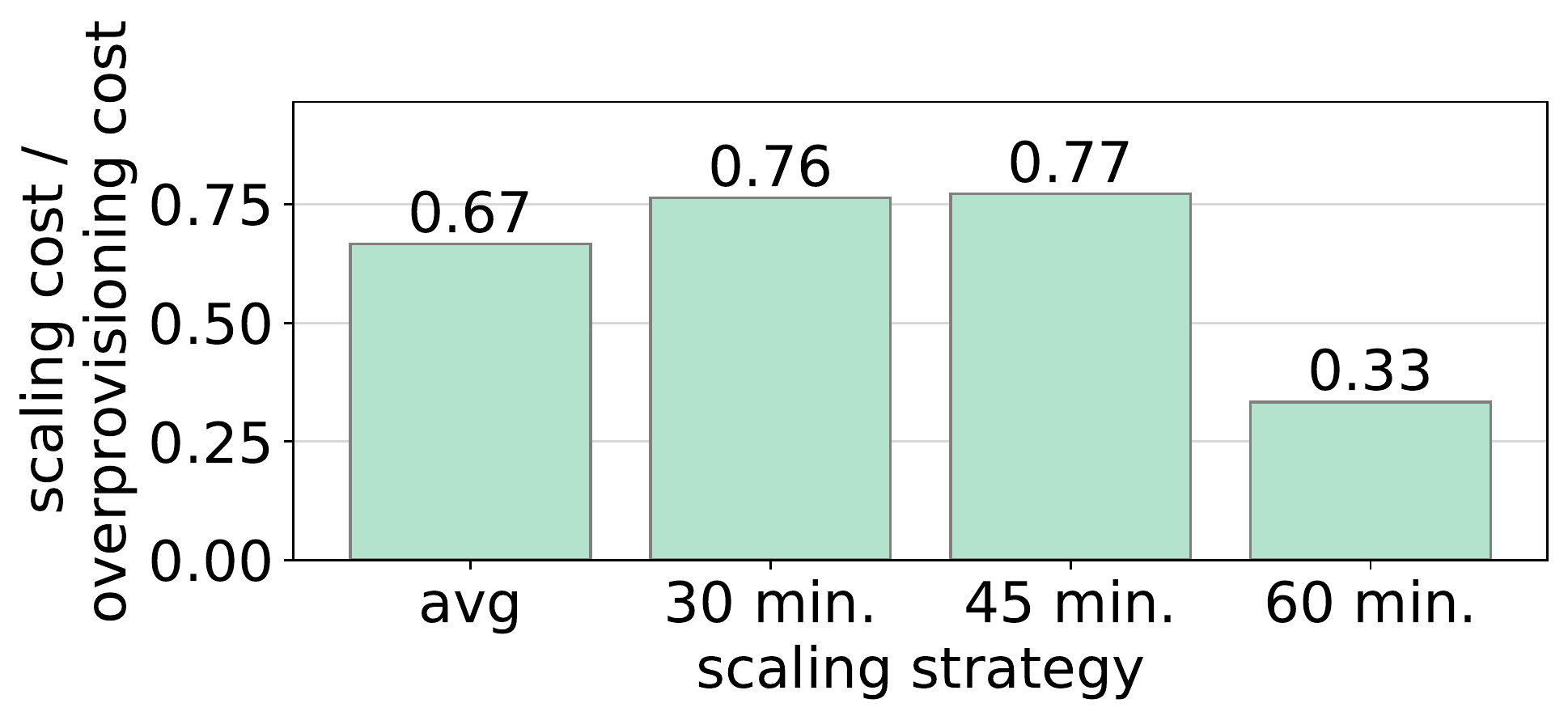}
        \label{fig:cost-hazard}%
    }\\
    \subfloat[][Remote driving scaling delay violations]{%
        \includegraphics[width=.3\textwidth]{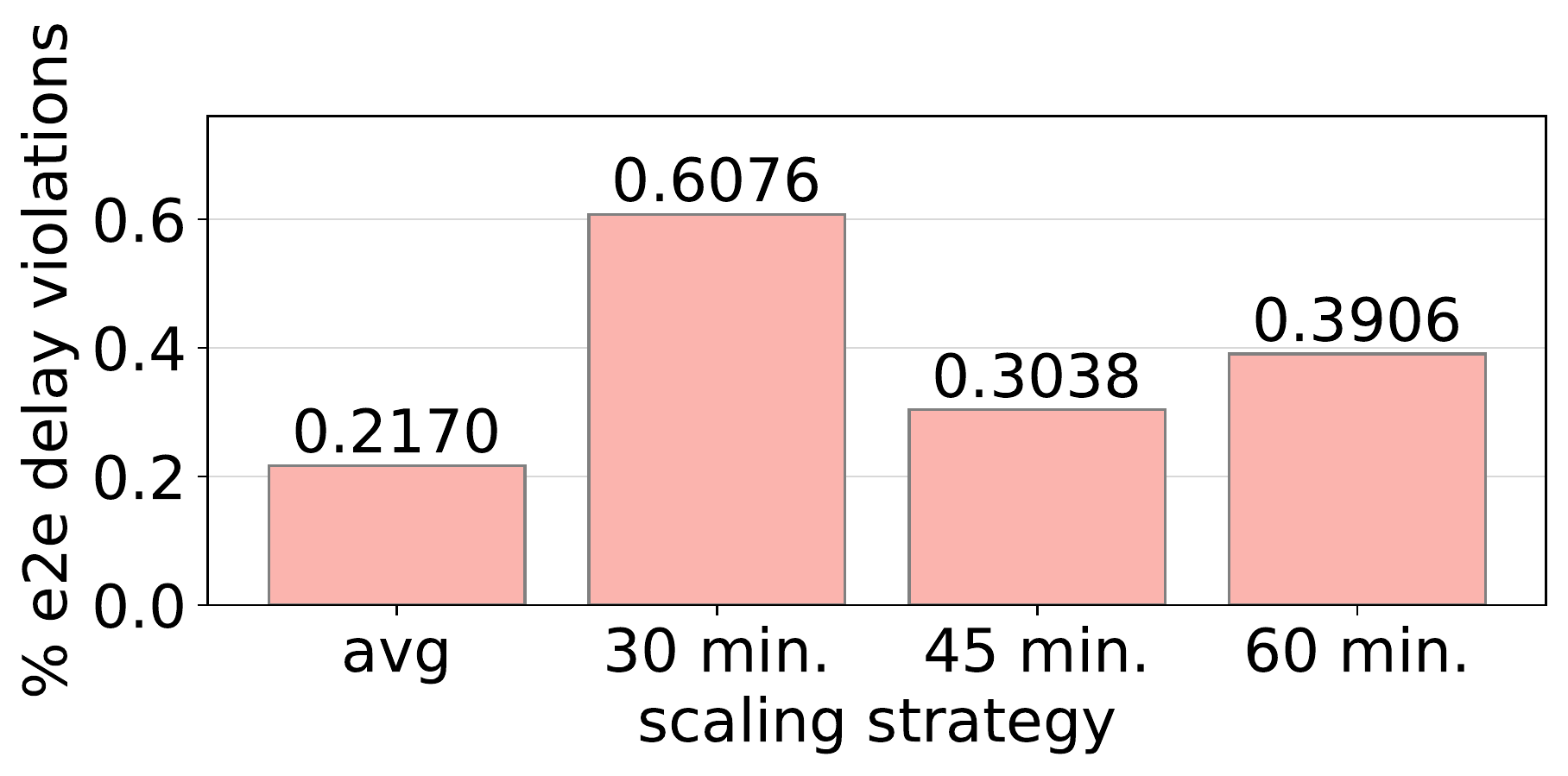}
        \label{fig:delay-remote}%
    }%
    \subfloat[][Cooperative awareness delay violations]{%
        \includegraphics[width=.3\textwidth]{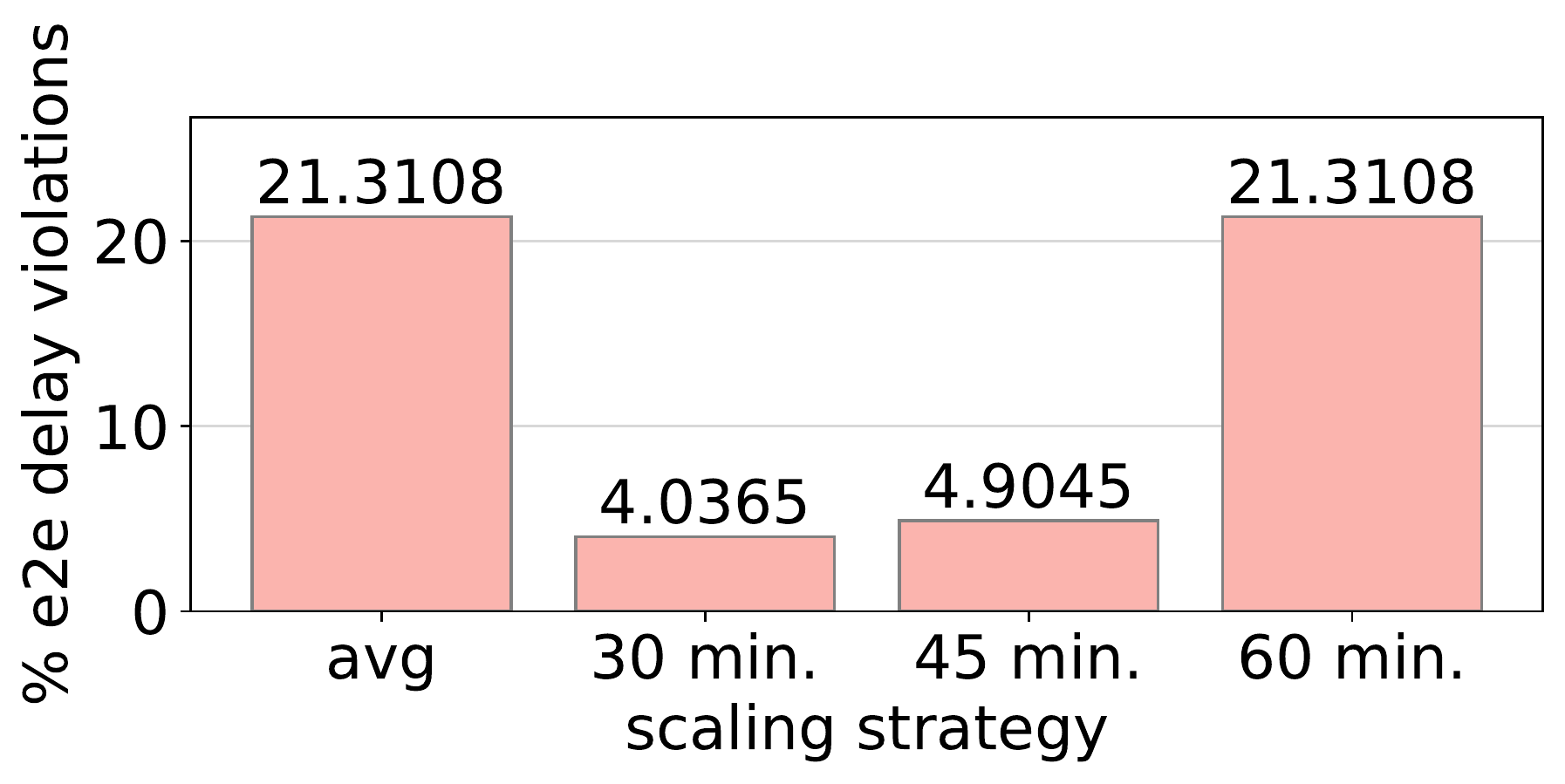}
        \label{fig:delay-cooperative}%
    }
    \subfloat[][Hazard warning delay violations]{%
        \includegraphics[width=.3\textwidth]{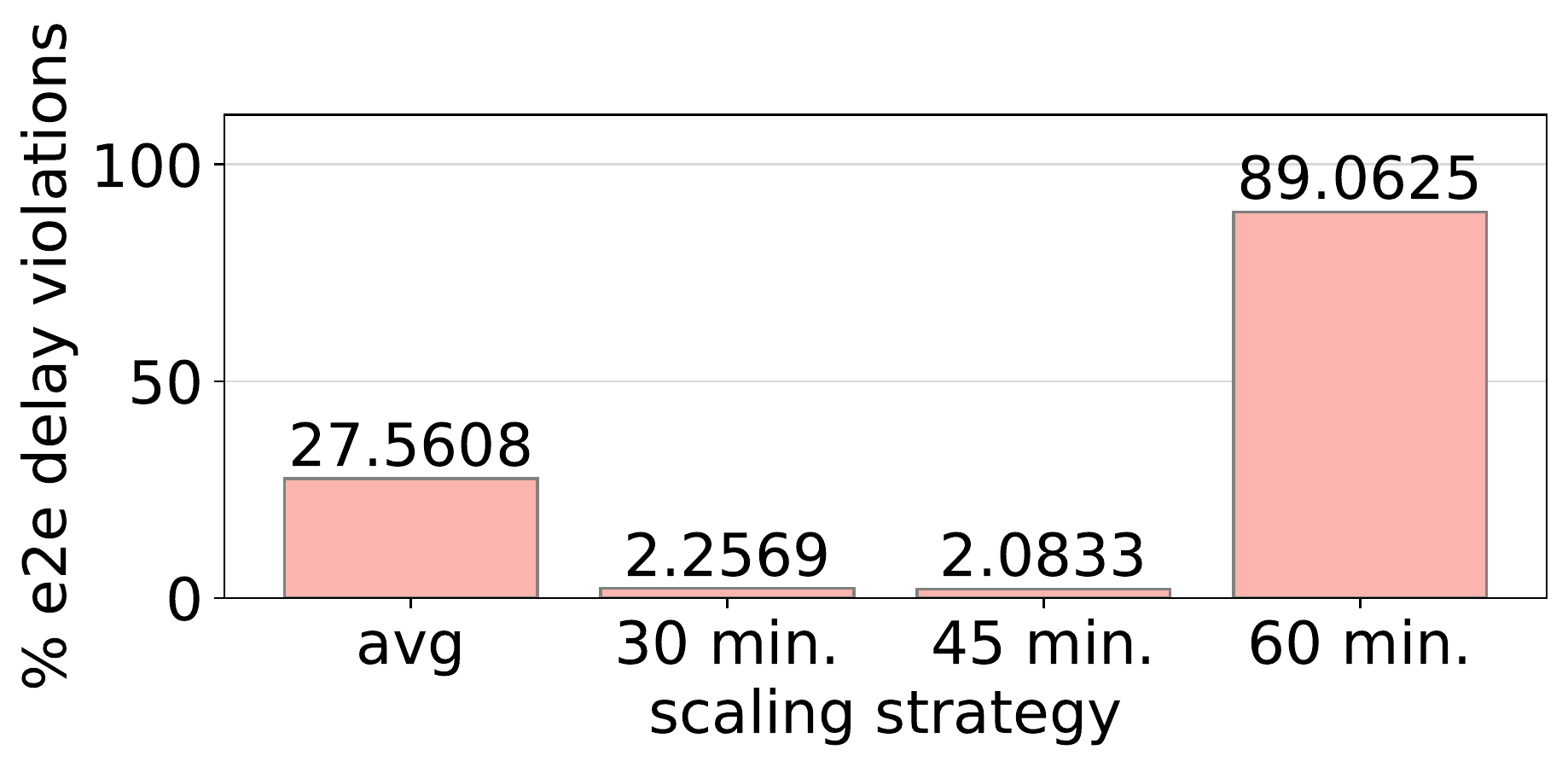}
        \label{fig:delay-hazard}%
    }
    \caption[]{
        Cost savings and delay violations due to scaling.
        TES with online training was used for n-min. strategies
        (i.e., Algorithm~\ref{alg:n-min-scaling}).
    }%
    \label{fig:scaling}%
\end{figure*}

\begin{figure}[h!]%
    \centering
    \subfloat[][]{%
        \includegraphics[width=0.4\textwidth]{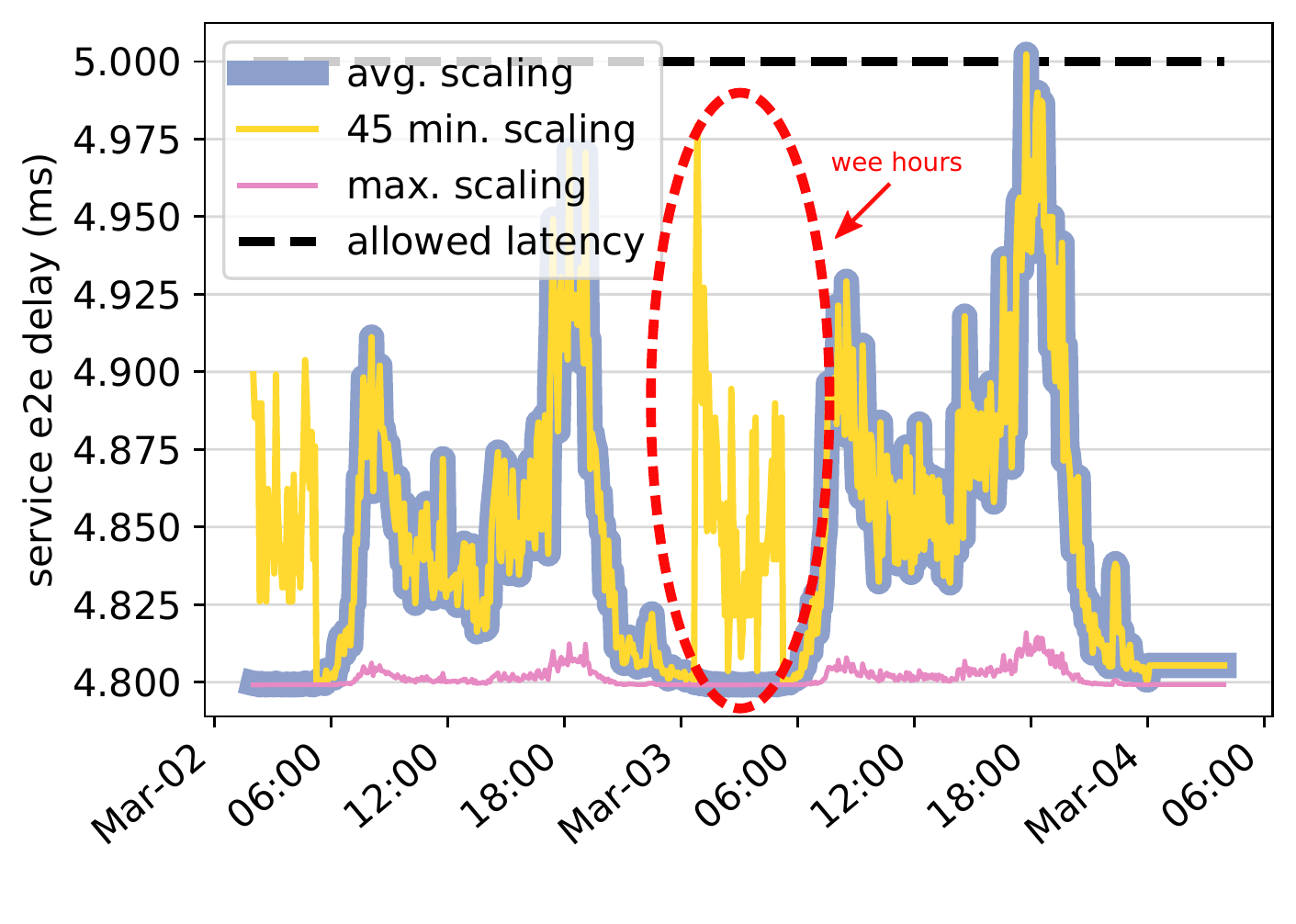}
        \label{fig:E2E-delay-evolution}%
    }\\
    \subfloat[][]{%
        \includegraphics[width=0.4\textwidth]{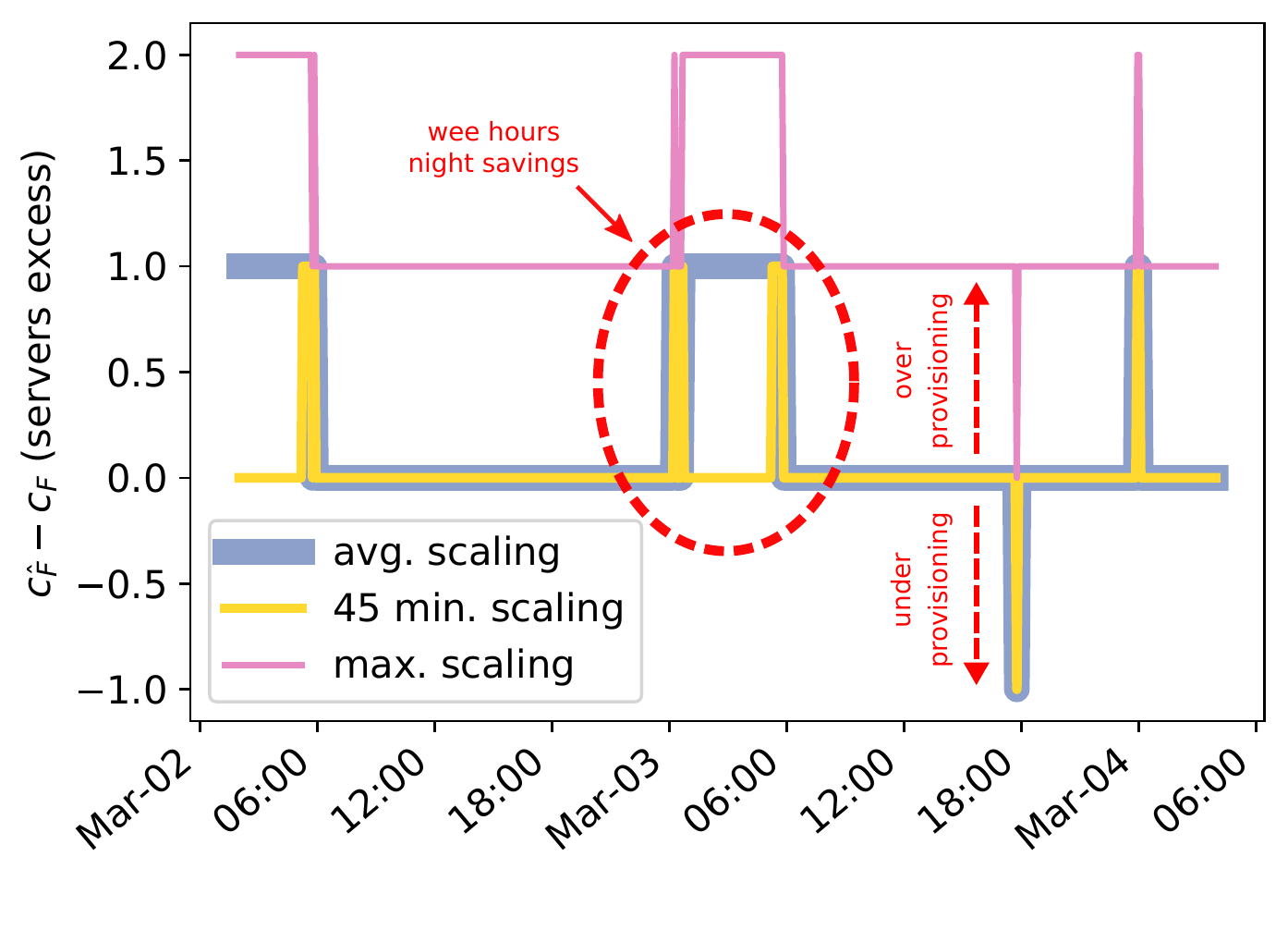}
        \label{fig:provisioning-comparison}%
    }
    \caption[]{
        Impact of remote driving scaling on
        (a) delay violations, and (b) difference of
        forecasted $c_{\hat{F}}$
        and required $c_F$ servers.
        TES with online training was used for 45-min. scaling
        (i.e., Algorithm~\ref{alg:n-min-scaling}).
        }
    \label{fig:remote-driving-comparisons}%
\end{figure}



Figure~\ref{fig:scaling} compares the performance of
\textit{over-provisioning-scaling}, \textit{avg.~scaling}, and \textit{$n$-min~scaling}
in the remote driving
(see Figures~\ref{fig:scaling}a and \ref{fig:scaling}d),
cooperative awareness
(see Figures~\ref{fig:scaling}b and \ref{fig:scaling}e),
and hazard warning (see Figures~\ref{fig:scaling}c and \ref{fig:scaling}f)
V2N services.
The three scaling strategies were tested under simulation
in the \textit{non-COVID-19}
scenario, as the traffic flow was significantly higher than
in the \textit{COVID-19}.
In every simulation $n$ was set to 30, 45, and 60 minutes for
the \textit{$n$-min~scaling} strategy, assuming scaling operations 
take less than 30 minutes.
Figures~\ref{fig:scaling}a-\ref{fig:scaling}c compare the cost of
\textit{avg.~scaling} and \textit{$n$-min~scaling} against 
\textit{over-provisioning scaling};
and Figures~\ref{fig:scaling}d-\ref{fig:scaling}f report the ratio of E2E violations.

For the remote driving simulations in Figures~\ref{fig:scaling}a and \ref{fig:scaling}d,
the service rate, and target latency were set to
$(\mu=\mu_{EVS}, T_0=5ms)$.
Results show that \textit{n-min.} reduces the costs with respect
to both an \textit{over-provisioning} strategy, and the \textit{avg. scaling}.
These savings are attained during the night, when the vehicular traffic
on the streets drop and it is no longer necessary to have that many
computing servers $c$ to process the traffic.
Figure~\ref{fig:E2E-delay-evolution} and
Figure~\ref{fig:provisioning-comparison} depict the service E2E delay
a excess of servers using the different scaling strategies. The night
saving are appreciated in the wee hours of the morning of March~3rd
(see Figure~\ref{fig:provisioning-comparison}),
when the \textit{45 min. scaling} decreases by one the number of servers
given the drop of traffic, whilst the \textit{avg. scaling} keeps the
same number of active servers, which results in a resource over 
provisioning leading to higher costs. Moreover, even though the
\textit{45 min. scaling} decreased the number of servers in the first
hours of March~3rd, Figure~\ref{fig:E2E-delay-evolution} shows that
still the service E2E delay remained below the 5~ms latency constraint.
Although the reader might think that the \textit{n-min. scaling}
strategies might substantially increase the percentage of
E2E delay violations, Figure~\ref{fig:delay-remote} shows that
at most, there is only an increase of $\le 0.4\%$ of E2E
delay violations over the simulated period. Additionally,
such delay violations are assumable noticing the
scaling savings, which go up to a $5\%$ according to
Figure~\ref{fig:cost-remote}.


In the cooperative awareness simulations' of
Figure~\ref{fig:scaling}b and \ref{fig:scaling}e, the service rate and target latency
were set to $(\mu=\mu_{EVS}/20, T_0=100ms)$.
Given the target latency
$T_0=100ms$,
throughout the experiments it was enough to deploy only
$c=1$ instances of the service almost every
time. Thus, both the \textit{average scaling}, and
\textit{over-provisioning scaling} strategies cost
the same (ratio equal to 1 in Figure~\ref{fig:scaling}b).
In the case of the \textit{$n$-min. scaling} strategy,
setting up $n$ to 30 and 45 minutes
result into higher costs than the
\textit{over-provisioning scaling} (ratio above 1
on the left axis in Figure~\ref{fig:scaling}b),
as both setups over-estimate the required resources.
Consequently, \textit{$n$-min. scaling} leads to less E2E delay
violations than \textit{avg. scaling} (above 4 times less,
$\frac{21.3108~\%}{4.9045~\%}=4.34$ to be specific).

For the last V2N service, the hazard warning, every
simulation used a service rate and target latency
of $(\mu=\mu_{EVS}/2, T_0=10ms)$.
Figure~\ref{fig:scaling}f show that
\textit{$n$-min. scaling} with $n=30$ and $n=45$
achieve around 12 times
($\frac{27.5608~\%}{2.2569~\%} = 12.21$)
less E2E delay violations than the 
\textit{avg. scaling} strategy.
The reduction of violations only incur in, at maximum,
less than a 15~\% ($\tfrac{0.77}{0.67}=14.92~\%$ of additional
investment over the \textit{avg.~scaling} strategy.
However, \textit{$n$-min. scaling}
underestimates the required resources when $n=60$, and it incurs
into more E2E delay violations than \textit{avg. scaling}.

\section{Conclusions and future work}
\label{sec:conclusions}
This paper provides an extensive analysis of
state-of-the-art techniques to forecast the road traffic
of Torino city, either leveraging on time-series or ML-based techniques.
The performed analysis compares each forecasting technique's RMSE considering
\textit{(i)} forecasting intervals from 5 to 60 minutes,
\textit{(ii)} offline/online training;
\textit{(iii)} COVID-19 lockdown; and
\textit{(iv)} neighboring road probes.
Results show that under offline training, ML-based techniques outperform traditional
time-series methods, especially during the COVID-19
lockdown, as they adapted to the Torino traffic
drop.
Whilst with online training, time-series
techniques achieve results better or as good as
the analyzed ML-based techniques.
However, none of the analyzed methods could
benefit from information of neighboring stations.
%
Experimental results confirm the benefits of applying
three different forecast-based scaling solutions
to dimension of V2N services.
Savings of up to a 5\% only incur in an increase of
$\le 0,4\%$ of latency violations in the remote driving
use case.
For the cooperative awareness
an extra 31\% of investment
achieved a 4-fold latency
reduction,
whilst for the hazard
warning less than a 15\%
investment increase
already
resulted in a 12-fold
latency reduction.


A first direction to extend this work is to find techniques that can incorporate
neighboring road probes' information, such as spatial analysis techniques.
Furthermore, the applicability of the presented techniques to different scenarios is also envisioned as a next step of this work. The use of different datasets, including operator records with respect to the base stations used by mobile phones to access the Internet, is going to be taken into consideration. In such scenario, forecasting the user density distribution along time would enable better decisions regarding the edge server placement and service migrations. 

Similarly, to the adopted scaling strategy of this work, enhancing orchestration algorithms with forecasting information would contribute to a smarter orchestration and resource control. Resulting decisions would be impacted in terms of improved quality, accuracy, and optimality. Optimized deployment, enhanced management and control of elastic network slices that support dynamic demands and their respective SLAs, improved resource arbitration and allocation and maximized service request admission are some examples where forecasting information can impact the decisions.

The aforementioned mechanisms are going to be developed and leveraged in selected use cases in the scope of the 5Growth project, which comprises Industry 4.0, transportation and energy scenarios, targeting full support of automation and SLA control for vertical services life-cycle management.
Hence, it would be worth-studying the probability of
forecasting less demand than what is required by each use case,
i.e., $\mathbb{P}(\hat{F}<F)$; so as to perform preemptive
actions under high probabilities of forecasting error.
Such a calculus deserves a detailed analysis on how
to compute $\max$-statistics for
correlated random variables (e.g. speed and traffic flow) \cite{maxjointprob}.



\section*{Acknowledgment}

This work has been partially funded by the EC H2020 5GROWTH Project (grant no. 856709), H2020 collaborative Europe/Taiwan research project 5G-DIVE (grant no.~859881),
and ``Excelencia para el Profesorado Universitario'' program of Madrid regional
government.

\bibliographystyle{IEEEtran}
\bibliography{main}

%









\end{document}